\documentclass[useAMS,usenatbib,usedcolumn,usegraphicx]{mn2e}
\usepackage{times}
\usepackage{lscape}
\def\teff{\ensuremath{T_{\mathrm{eff}}}}
\def\logg{log~\ensuremath{g}}
\def\Logg{Log~\ensuremath{g}}

\def\h2{H\ensuremath{_\mathrm{2}}}

\def\heabund{\ensuremath{\mathrm{\frac{He}{H}}}}

\def\chis{\ensuremath{\chi^2}}
\def\pom{\ensuremath{\pm}}
\newcommand{\lnum}[2]{#1\,$\times$\,10$^{#2}$}
\newcommand{\spec}[2]{#1\,\textsc{#2}}

\title[Heavy element abundances in DAOs]
  {Heavy element abundances in DAO white dwarfs measured from \textit{FUSE}\ data}
\author[S.A. Good et al.]
  {S.A. Good$^1$\thanks{Email: sag15@le.ac.uk},
  M.A. Barstow$^1$, M.R. Burleigh$^1$, P.D. Dobbie$^1$, J.B. Holberg$^2$ and \newauthor I. Hubeny$^3$ \\
  $^1$Department of Physics and Astronomy, University of Leicester, University Road, Leicester LE1 7RH \\
  $^2$Lunar and Planetary Laboratory, University of Arizona, Tucson, AZ 85721, USA \\
  $^3$Steward Observatory and Department of Astronomy, University of Arizona, Tucson, AZ 85721, USA}
\date{Released 2002 Xxxxx XX}

\pagerange{\pageref{firstpage}--\pageref{lastpage}} \pubyear{2004}

\def\LaTeX{L\kern-.36em\raise.3ex\hbox{a}\kern-.15em
    T\kern-.1667em\lower.7ex\hbox{E}\kern-.125emX}

\begin{document}

\label{firstpage}

\maketitle

\begin{abstract}
We present heavy element abundance measurements for 16 DAO white dwarfs, determined from \textit{Far-Ultraviolet Spectroscopic Explorer}\ (\textit{FUSE}) spectra.  Evidence of absorption by heavy elements was found in the spectra of all the objects.  Measurements were made using models that adopted the temperatures, gravities and helium abundances determined from both optical and \textit{FUSE}\ data by \citet{good04}.  It was found that, when using the values for those parameters measured from optical data, the carbon abundance measurements follow and extend a similar trend of increasing abundance with temperature for DA white dwarfs, discovered by \citet{bars03abund}.  However, when the \textit{FUSE}\ measurements are used the DAO abundances no longer join this trend since the temperatures are higher than the optical measures.  Silicon abundances were found to increase with temperature, but no similar trend was identified in the nitrogen, oxygen, iron or nickel abundances, and no dependence on gravity or helium abundances were noted.  However, the models were not able to reproduce the observed silicon and iron line strengths satisfactorily in the spectra of half the objects, and the oxygen features of all but three.  Despite the different evolutionary paths that the types of DAO white dwarfs are thought to evolve through, their abundances were not found to vary significantly, apart from for the silicon abundances.

Abundances measured when the \textit{FUSE}\ derived values of temperature, gravity and helium abundance were adopted were, in general, a factor 1-10 higher than those determined when the optical measure of those parameters was used.  Satisfactory fits to the absorption lines were achieved in approximately equal number.  The models that used the \textit{FUSE}\ determined parameters seemed better at reproducing the strength of the nitrogen and iron lines, while for oxygen, the optical parameters were better.  For the three objects whose temperature measured from \textit{FUSE}\ data exceeds 120\,000 K, the carbon, nitrogen and oxygen lines were too weak in the models that used the \textsc{FUSE}\ parameters.  However, the model that used the optical parameters also did not reproduce the strength of all the lines accurately.  
\end{abstract}


\section{Introduction}

DAO white dwarfs, for which the prototype is HZ\,34 \citep{koes79,wese93}, are
characterised by the presence of He\,\textsc{ii}\ absorption in their optical
spectra in addition to the hydrogen Balmer series.  Radiative forces cannot
support sufficient helium in the line forming region of the white dwarf to
reproduce the observed lines \citep{venn88}, and one explanation for their
existence is that they are transitional objects switching between the helium-
and hydrogen-rich cooling sequences \citep{font87}.  If a small amount of
hydrogen were mixed into an otherwise helium dominated atmosphere, gravitational settling would then create a thin hydrogen layer at the surface
of the white dwarf, with the boundary between the hydrogen and helium described
by diffusive equilibrium.  However, \citet{napi93}\ found that the
line profile of the \spec{He}{ii}\ line at 4686 \AA\ in the DAO S\,216 was better matched by homogeneous composition models, rather than the predicted layered configuration.   Subsequently, a spectroscopic investigation
by \citet{berg94}\ found that the \spec{He}{ii}\ line
profile of only one out of a total of 14 objects was better reproduced by
stratified models.  In addition, the line profile of one object (PG\,1210+533)
could not be reproduced satisfactorily by either set of models.

Most of the DAOs analysed by \citeauthor{berg94}\ were comparatively hot for
white dwarfs, but with low gravity, which implies that they have low mass. 
Therefore, they may have not have been massive enough to ascend the asymptotic
giant branch, and instead may have evolved from the extended horizontal
branch.  \citet{berg94}\ suggested that a process such as weak mass loss may be
occurring in these stars, which might support the observed quantities of helium
in the line forming regions of the DAOs \citep{ungl98,ungl00}.  Three of the
\citet{berg94}\ objects (RE\,1016-053, PG\,1413+015 and RE\,2013+400) had
comparatively `normal' temperatures and gravities, yet helium absorption
features were still observed.  Each of these are in close binary systems with M
dwarf (dM) companions.  It may be that as the white dwarf progenitor passes
through the common envelope phase, mass is lost, leading to the star being
hydrogen poor.  Then, a process such as weak mass loss could mix helium into
the line forming region of the white dwarf.  Alternatively, these DAOs might be
accreting from the wind of their companions, as is believed to be the case for
another DAO+dM binary, RE\,0720-318 \citep{dobb99}.

Knowledge of the effective temperature (\teff) and surface gravity (g) of a
white dwarf is vital to our understanding of its evolutionary status.  Values
for both these parameters can be found by comparing the profiles of the
observed hydrogen Balmer lines to theoretical models.  This technique was
pioneered by \citet{holb85}\ and extended to a large sample of white dwarfs by
\citet{berg92}.  However, for objects in close binary systems, where the white
dwarf cannot be spatially resolved, the Balmer line profiles cannot be used as
they are frequently contaminated by flux from the secondary (if it is of type K
or earlier). Instead, the same technique can be applied to the Lyman lines that
are found in far-ultraviolet (far-UV) data, as the white dwarf is much brighter
in this wavelength region than the companion \citep[e.g.][]{bars94sirius}. 
However, \citet{bars01}, \citet{bars03}\ and \citet{good04}\ have compared the
results of fitting the Balmer and Lyman lines of DA and DAO white dwarfs, and
found that above 50\,000 K, the \teff\ measurements begin to diverge.  This
effect was stronger in some stars; in particular, the Lyman lines of 3 DAOs in
the sample of \citet{good04}\ were so weak that the temperature of the best
fitting model exceeded 120\,000 K, which was the limit of their model grid.  

One factor that influences the measurements of temperature and gravity is the treatment of heavy element contaminants in the atmosphere of a white dwarf.   \citet{bars98}\ found that heavy element line blanketing significantly affected the Balmer line profiles in their theoretical models.  The result was a decrease in the measured \teff\ of a white dwarf compared to when a pure hydrogen model was used.  In addition, \citet{bars03abund}\ conducted a systematic set of measurements of the abundances of heavy elements in the atmospheres of hot DA white dwarfs, which differ from the DAOs in that no helium is observed.  They found that the presence or lack of heavy elements in the photosphere of the white dwarfs largely reflected the predictions of radiative levitation, although the abundances did not match the expected values very well.

We have performed systematic measurements of the heavy element abundances in \textit{Far-Ultraviolet Spectroscopic Explorer}\ (\textit{FUSE}) observations of DAO white dwarfs.  The motivation for this work was twofold: firstly, we wish to investigate if the different evolutionary paths suggested for the DAOs are reflected in their heavy element abundances, as compared to the DAs of \citet{bars03abund}.  Secondly, since \teff\ and \logg\ measurements from both optical and far-UV data for all the DAOs in our sample have previously been published \citep{good04}, we investigate which set of models better reproduces the strengths of the observed lines.  The paper is organised as follows: in \S\ref{observations}, \S\ref{models}\ and \S\ref{dataanalysis}\ we describe the observations, models and data analysis technique used.  Then, in \S\ref{results}\ the results of the abundance measurements are shown.  The ability of the models to reproduce the observations are discussed in \S\ref{discussion}\ and finally the conclusions are presented in \S\ref{conclusions}.

\section{Observations}
\label{observations}

\begin{figure}
	\includegraphics[]{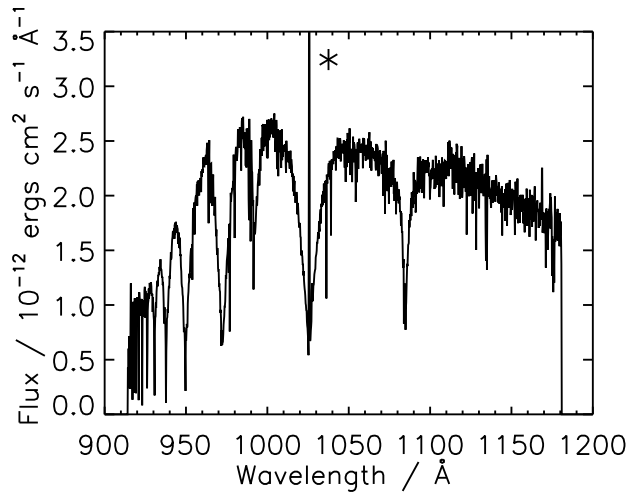}
	\caption{\label{pg1210spectrum} \textit{FUSE}\ spectrum of PG\,1210+533, produced by combining the spectra from the individual detector segments.  * The emission in the core of the Lyman beta H\,\textsc{i}\ line is due to terrestrial airglow.}
\end{figure}

Far-UV data for all the objects were obtained by the \textit{FUSE}\
spectrographs and cover the full Lyman series, apart from Lyman $\alpha$. 
Table \ref{fuseobs}\ summarizes the observations, which were downloaded by us
from the Multimission Archive (http://archive.stsci.edu/mast.html), hosted by
the Space Telescope Science Institute.  Overviews of the \textit{FUSE}\ mission and in-orbit performance can be found in \citet{moos00} and \citet{sahn00} respectively.  Full details of our data extraction, calibration and co-addition techniques are published in \citet{good04}.  In brief, the data are calibrated using the \textsc{calfuse}\ pipeline version 2.0.5 or later, resulting in eight spectra (covering different wavelength segments) per \textit{FUSE}\ exposure.  The exposures for each segment are then co-added, weighting each according to exposure time, and finally the segments are combined, weighted by their signal-to-noise, to produce a single spectrum.  Before each co-addition, the spectra are cross-correlated and shifted to correct for any wavelength drift.  Figure \ref{pg1210spectrum}\ shows an example of the output from this process, for PG\,1210+533.

\begin{table*}
 \begin{center}
 \caption{List of \textit{FUSE} observations for the stars in the sample.  All observations used TTAG mode.}
 \label{fuseobs}
 \begin{tabular}{llcccc}
 \hline
 Object & WD number & Obs. ID & Date & Exp. Time (s) & Aperture \\
 \hline
 PN\,A66\,7     & WD0500-156 & B0520901 & 2001/10/05 & 11525 & LWRS \\
 HS\,0505+0112  & WD0505+012 & B0530301 & 2001/01/02 & 7303 & LWRS \\ 
 PN\,PuWe\,1    & WD0615+556 & B0520701 & 2001/01/11 & 6479 & LWRS \\
                &            & S6012201 & 2002/02/15 & 8194 & LWRS \\
 RE\,0720-318   & WD0718-316 & B0510101 & 2001/11/13 & 17723 & LWRS \\
 TON\,320       & WD0823+317 & B0530201 & 2001/02/21 & 9378 & LWRS \\
 PG\,0834+500   & WD0834+501 & B0530401 & 2001/11/04 & 8434 & LWRS \\
 PN\,A66\,31    &            & B0521001 & 2001/04/25 & 8434 & LWRS \\
 HS\,1136+6646  & WD1136+667 & B0530801 & 2001/01/12 & 6217 & LWRS \\
                &            & S6010601 & 2001/01/29 & 7879 & LWRS \\
 Feige\,55      & WD1202+608 & P1042105 & 1999/12/29 & 19638 & MDRS \\
 		&            & P1042101 & 2000/02/26 & 13763 & MDRS \\
                &            & S6010101 & 2002/01/28 & 10486 & LWRS \\
                &            & S6010102 & 2002/03/31 & 11907 & LWRS \\
                &            & S6010103 & 2002/04/01 & 11957 & LWRS \\
                &            & S6010104 & 2002/04/01 & 12019 & LWRS \\
 PG\,1210+533   & WD1210+533 & B0530601 & 2001/01/13 & 4731 & LWRS \\
 LB\,2          & WD1214+267 & B0530501 & 2002/02/14 & 9197 & LWRS \\
 HZ\,34         & WD1253+378 & B0530101 & 2003/01/16 & 7593 & LWRS \\
 PN\,A66\,39    &            & B0520301 & 2001/07/26 & 6879 & LWRS \\
 RE\,2013+400   & WD2011+395 & P2040401 & 2000/11/10 & 11483 & LWRS \\
 PN\,DeHt\,5    & WD2218+706 & A0341601 & 2000/08/15 & 6055 & LWRS \\
 GD\,561        & WD2342+806 & B0520401 & 2001/09/08 & 5365 & LWRS \\
 \hline
 \end{tabular}
 \end{center}
\end{table*}

\section{Model calculations}
\label{models}

A new grid of stellar model atmospheres was created using the non-local
thermodynamic (non-LTE) code \textsc{tlusty}\ (v. 195) \citep{hube95}\ and its
associated spectral synthesis code \textsc{synspec}\ (v. 45), based on the
models created by \citet{bars03abund}; as with their models, all calculations
were performed in non-LTE with full line-blanketing, and with a treatment of
Stark broadening in the structure calculation.  Grid points with varying
abundances were calculated for 8 different temperatures between 40\,000 and
120\,000 K, in 10\,000 K steps, 4 different values of \logg: 6.5, 7.0, 7.5 and
8.0, and, since the objects in question are DAOs, 5 values of log
$\frac{He}{H}$\ between -5 and -1, in steps of 1.  Therefore, the grid
encompassed the full range of the stellar parameters determined for the DAOs by
\citet{good04}.  Models were calculated for a range of heavy element
abundances, which were factors of 10$^{-1}$ to 10$^{1.5}$\ times the values
found for the well studied white dwarf G\,191-B2B in an earlier analysis \citep{bars01g191}\ (C/H =
\lnum{4.0}{-7}, N/H = \lnum{1.6}{-7}, O/H = \lnum{9.6}{-7}, Si/H =
\lnum{3.0}{-7}, Fe/H = \lnum{1.0}{-5}\ and Ni/H = \lnum{5.0}{-7}), with each
point a factor of 10$^{0.5}$\ different from the adjacent point.  Above this
value it was found that the \textsc{tlusty}\ models did not converge.  Since,
even with these abundances, the observed strengths of some lines could not be
reproduced, the range was further extended upwards by a factor of 10, using
\textsc{synspec}.  However, as scaling abundances in this way is only valid if
it can be assumed that the structure of the white dwarf atmosphere will not be
significantly affected by the change, abundance values this high should be
treated with caution. 

Molecular hydrogen is observed in the spectra of some the objects.  For those stars, the templates of \citet{mcca03}\ were used with the measurements of \citet{good04}\ to create appropriate molecular hydrogen absorption spectra.

\section{Data analysis}
\label{dataanalysis}

The measurements of heavy element abundances were performed using the spectral
fitting program \textsc{xspec}\ \citep{arna96}, which adopts a $\chi^2$\
minimisation techique to determine the set of model parameters that best
matches a data set.  As the full grid of synthetic spectra was large, it was
split into 4 smaller grids, one for each value of \logg.  The \textit{FUSE}\
spectrum for each white dwarf was split into sections, each containing
absorption lines due to a single element.  Table \ref{linelist}\ lists the
wavelength ranges used and the main species in those regions.  Where absorption
due to another element could not be avoided, the affected region was set to be ignored by the fitting process.  The appropriate model grid for the
object's \logg\ was selected, from the measurements of \citet{good04}, and the
\teff\ and helium abundance were constrained to fall within the 1$\sigma$\
limits from the same measurements, but were allowed to vary freely within those
ranges.  For those stars whose spectra exhibit molecular hydrogen absorption, the molecular hydrogen model created for that object was also included.  Then, the model abundance that best matched the absorption features in
the real data was found, and 3$\sigma$\ errors calculated from the
$\Delta\chi^2$\ distribution.  Where an element was not detected, the
3$\sigma$\ upper bound was calculated instead.  This analysis was repeated
separately for the parameters determined by \citet{good04}\ from both optical
and \textit{FUSE}\ data -- these parameters are listed in Table
\ref{fuseoptresults}. 

\begin{table}
\begin{center}
  \caption{Wavelength regions chosen for spectral fitting, the main species in each range and the central wavelengths of strong lines.}
  \label{linelist}  
  \begin{tabular}{lr@{\,--\,}lc}
  \hline
  Species & \multicolumn{2}{c}{Wavelength range / \AA} & Central wavelengths / \AA\\
  \hline
  C\,\textsc{iv}   & 1107.0 & 1108.2 & 1107.6, 1107.9, 1108.0 \\
                   & 1168.5 & 1170.0 & 1168.8, 1169.0 \\
  N\,\textsc{iii}  & 990.5  & 992.5  & 991.6 \\
  N\,\textsc{iv}   & 920.0  & 926.0  & 992.0, 922.5, 923.1, \\
                   & \multicolumn{2}{c}{}  & 923.2, 923.7, 924.3 \\
  O\,\textsc{iv}   & 1067.0 & 1069.0 & 1067.8 \\
  O\,\textsc{vi}   & 1031.0 & 1033.0 & 1031.9 \\
                   & 1036.5 & 1038.5 & 1037.6 \\
  Si\,\textsc{iii} & 1113.0 & 1114.0 & 1113.2 \\
  Si\,\textsc{iv}  & 1066.0 & 1067.0 & 1066.6, 1066.7 \\
                   & 1122.0 & 1129.0 & 1122.5, 1128.3 \\
  Fe\,\textsc{v}   & 1113.5 & 1115.5 & 1114.1 \\
  Fe\,\textsc{vi/vii}  & 1165.0 & 1167.0 & 1165.1, 1165.7,1166.2\\
  Fe\,\textsc{vii} & 1116.5 & 1118.0 & 1117.6 \\
  Ni\,\textsc{v}   & 1178.0 & 1180.0 & 1178.9 \\
  \hline
  \end{tabular}
 \end{center}
\end{table}

\begin{table*}
  \begin{center}
    \caption{Mean best fitting parameters obtained from optical and FUSE data, from \citet{good04}.     Log~\heabund\ is not
    listed for the optical data of PG\,1210+533 as it is seen to vary between
    observations.  1$\sigma$ confidence intervals are given, apart from where
    the temperature or gravity of the object is beyond the range of
    the model grid and hence a \chis\ minimum has not been reached.  Where a parameter is beyond the range of the model grid, the value is written in italics.  The final columns show the total molecular hydrogen column density for each star, where it was detected, and the measured interstellar extinction, also from \citet{good04}.}
    \label{fuseoptresults}
    \begin{tabular}{lcccccccc}
      \hline
      & \multicolumn{3}{l}{\textbf{BALMER LINE SPECTRA}} & \multicolumn{5}{l}{\textbf{LYMAN LINE SPECTRA}} \\
      Object & \teff\ / K & \Logg & Log~\heabund
      & \teff\ / K & \Logg & Log~\heabund & Log~\h2 & E(B-V) \\
      \hline
      PN\,A66\,7 & 66955 \pom 3770 & 7.23 \pom 0.17 & -1.29 \pom 0.15 & 99227 \pom 2296 & 7.68 \pom 0.06 & -1.70 \pom 0.09 & 18.5 & 0.001 \\
      HS\,0505+0112 & 63227 \pom 2088 & 7.30 \pom 0.15 & \textit{-1.00} &
      \textit{120000} & 7.24 & \textit{-1.00} & & 0.047 \\
      PN\,PuWe\,1 & 74218 \pom 4829 & 7.02 \pom 0.20 & -2.39 \pom 0.37 & 109150 \pom
      11812 & 7.57 \pom 0.22 & -2.59 \pom 0.75 & 19.9 & 0.090 \\
      RE\,0720-318 & 54011 \pom 1596 & 7.68 \pom 0.13 & -2.61 \pom 0.19 & 54060
      \pom\ 776 & 7.84 \pom 0.03 & -4.71 \pom 0.69 & & 0.019 \\
      TON\,320 & 63735 \pom 2755 & 7.27 \pom 0.14 & -2.45 \pom 0.22 & 99007 \pom
      4027 & 7.26 \pom 0.07 & -2.00 \pom 0.12 & 14.9 & 0.000 \\
      PG\,0834+500 & 56470 \pom 1651 & 6.99 \pom 0.11 & -2.41 \pom 0.21 & \textit{120000} & 7.19 & \textit{-5.00} & 18.2 & 0.033 \\
      PN\,A66\,31 & 74726 \pom 5979 & 6.95 \pom 0.15 & -1.50 \pom 0.15 & 93887 \pom 3153 & 7.43 \pom 0.15 & \textit{-1.00} & 18.8 & 0.045 \\
      HS\,1136+6646 & 61787 \pom\ 700 & 7.34 \pom 0.07 & -2.46 \pom 0.08 &
      \textit{120000} & \textit{6.50} & \textit{-1.00} & & 0.001 \\
      Feige\,55 & 53948 \pom\ 671 & 6.95 \pom 0.07 & -2.72 \pom 0.15 & 77514 \pom\ 532 & 7.13 \pom 0.02 & -2.59 \pom 0.05 & & 0.023 \\
      PG\,1210+533 & 46338 \pom\ 647 & 7.80 \pom 0.07 & - & 46226 \pom\ 308 &
      7.79 \pom 0.05 & -1.03 \pom 0.08 & & 0.000 \\
      LB\,2 & 60294 \pom 2570 & 7.60 \pom 0.17 & -2.53 \pom 0.25 & 87622 \pom 3717 & 6.96 \pom 0.04 & -2.36 \pom 0.17 & & 0.004 \\
      HZ\,34 & 75693 \pom 5359 & 6.51 \pom 0.04 & -1.68 \pom 0.23 & 87004 \pom
      5185 & 6.57 \pom 0.20 & -1.73 \pom 0.13 & 14.3 & 0.000 \\
      PN\,A66\,39 & 72451 \pom 6129 & 6.76 \pom 0.16 & \textit{-1.00} & 87965 \pom 4701 & 7.06 \pom 0.15 & -1.40 \pom 0.14 & 19.9 & 0.130 \\
      RE\,2013+400 & 47610 \pom\ 933 & 7.90 \pom 0.10 & -2.80 \pom 0.18 & 50487
      \pom\ 575 & 7.93 \pom 0.02 & -4.02 \pom 0.51 & & 0.010 \\
      PN\,DeHt\,5 & 57493 \pom 1612 & 7.08 \pom 0.16 & -4.93 \pom 0.85 & 59851 \pom
      1611 & 6.75 \pom 0.10 & \textit{-5.00} & 20.1 & 0.160 \\
      GD\,561 & 64354 \pom 2909 & 6.94 \pom 0.16 & -2.86 \pom 0.35 & 75627 \pom
      4953 & 6.64 \pom 0.06 & -2.77 \pom 0.24 & 19.8 & 0.089 \\
      \hline
    \end{tabular}
  \end{center}
\end{table*}

\section{Results}
\label{results}

\begin{figure}
	\includegraphics[]{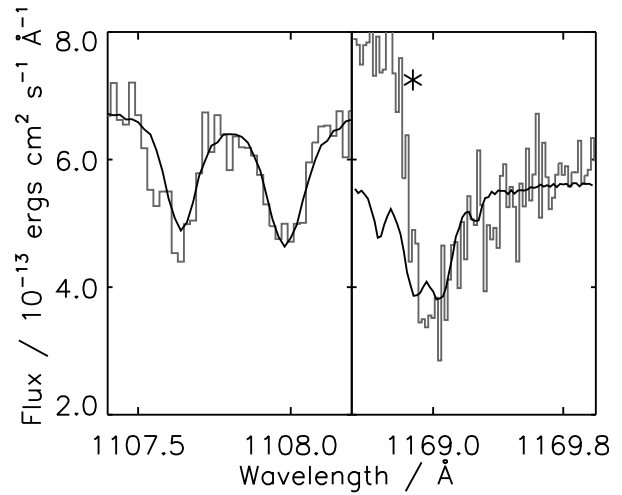}
	\caption{\label{lb2c} The C\,\textsc{iv}\ lines observed in the spectrum of LB\,2 (grey histogram), overlaid with the best fitting model (black line).  The temperature, gravity and helium abundance of the model were those measured from \textit{FUSE}\ data.  * Bins containing the high fluxes that can be seen on the blue side of the right panel and that are not predicted by the model were excluded from the fit.}
\end{figure}

\begin{figure}
	\includegraphics[]{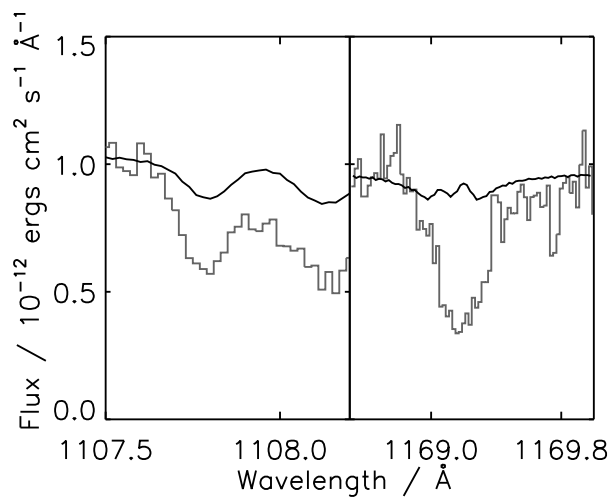}
	\caption{\label{hs0505c} The C\,\textsc{iv}\ lines observed in the spectrum of HS\,0505+0112 (grey histogram), overlaid with the best fitting model (black line).  The temperature, gravity and helium abundance of the model were those measured from \textit{FUSE}\ data.  The heavy element abundance of the model is the maximum possible for the model grid used in the fit, yet the predicted lines are much weaker than those observed.}
\end{figure}

\begin{figure}
	\includegraphics[]{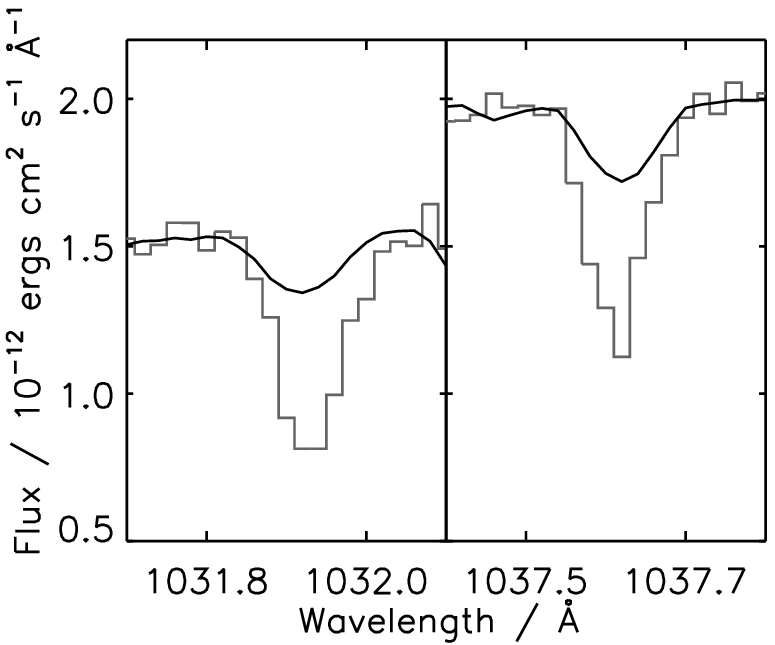}
	\caption{\label{re2013fuse_o} The O\,\textsc{vi}\ lines observed in the spectrum of RE\,2013+400 (grey histogram), overlaid with a model that has an oxygen abundance of 10 times the values measured for G\,191-B2B, and which uses the temperature, gravity and helium abundance measured from \textit{FUSE}\ data (black line).  The lines in the model are much weaker than those seen in the real data, no matter what abundance is chosen, and hence no oxygen abundance was recorded for this object.}
\end{figure}

The results of the analysis are listed in Table \ref{abundanceresults}, and a summary of which lines were detected in each spectrum is shown in Table \ref{detections}.  Heavy elements were detected in all objects studied.  Carbon and nitrogen were identified in all objects, although the strength of the lines could not be reproduced with the abundances included in the models in some cases.  Figure \ref{lb2c}\ shows an example of a fit to the carbon lines in the spectrum of LB\,2, where the model successfully reproduced the strength of the lines in the real data.  However, when a fit to the carbon lines in the spectrum of HS\,0505+0112 was attempted, it was found that, when the values of temperature, gravity and helium abundance determined from \textit{FUSE}\ data were used, the line strengths predicted by the model were far too weak compared to what is observed (Figure \ref{hs0505c}).  Oxygen lines were found to be, in general, particularly difficult to fit, with no abundance measurement recorded in a number of cases because the reproduction of the lines was so poor.  An example of a poor fit is shown in Figure \ref{re2013fuse_o}.  Silicon was also detected in all the spectra, but not iron.  Nickel abundances were very poorly constrained, with many non-detections, although the error margins spanned the entire abundance range of the model grid in three cases.   

\begin{table*}
    \caption{Measurements of abundances, with all values
    expressed as a number fraction with respect to hydrogen, and their 3$\sigma$\ confidence intervals.  Where the abundance value or the upper confidence boundary exceeded the highest abundance of the model grid, an $\infty$\ symbol is used.  Where an abundance or a lower error boundary is below the lowest abundance in the model grid, 0 is placed in the table; if this occurs for the abundance, the value in the +3$\sigma$\ column is an upper limit on the abundance.  In the table, * indicates that the model fit to the lines was poor, with $\chi^2_{red}$\ greater than 2, while \#\ is written instead of an abundance where the model was unable to recognisably reproduce the shape of the lines.}
    \label{abundanceresults}
    \begin{tabular}{lr@{$^+_-$}l@{$\times$10}lr@{$^+_-$}l@{$\times$10}lr@{$^+_-$}l@{$\times$10}lr@{$^+_-$}l@{$\times$10}l}
      \hline
      Object & \multicolumn{6}{c}{C/H} & \multicolumn{6}{c}{N/H} \\
      & \multicolumn{3}{c}{Optical} & \multicolumn{3}{c}{\textit{FUSE}} & \multicolumn{3}{c}{Optical} & \multicolumn{3}{c}{\textit{FUSE}} \\
      
      \hline

      PN\,A66\,7    & 1.38 & $_{0.531}^{1.54}$ & $^{-5}$ 
                    & 4.17 & $_{1.09}^{3.84}$ & $^{-5}$
		    & 7.96 & $_{5.96}^{12.2}$ & $^{-8}$
		    & 6.43 & $_{5.96}^{23.3}$ & $^{-6}$ \\       
			      
      H\,0505+0112  & 1.11 & $_{0.363}^{\infty}$ & $^{-4}$
                    & \multicolumn{3}{c}{$\infty$} 
		    & 7.36$^{*}$ & $_{4.60}^{7.98}$ & $^{-8}$
		    & \multicolumn{3}{c}{$\infty$} \\              			      
      PN\,PuWe\,1   & 1.44$^{*}$ & $_{0.975}^{2.22}$ & $^{-5}$
                    & 4.00$^{*}$ & $_{1.74}^{5.08}$ & $^{-5}$
		    & \multicolumn{3}{c}{$\infty$}
		    & \multicolumn{3}{c}{$\infty$} \\ 
                               
      RE\,0720-318  & 1.10 & $_{0.563}^{0.728}$ & $^{-6}$
                    & 2.97 & $_{1.45}^{1.90}$ & $^{-6}$
		    & 1.29 & $_{0.714}^{0.778}$ & $^{-7}$ 
                    & 1.21 & $_{0.467}^{0.689}$ & $^{-7}$ \\           
			      
      Ton\,320      & 7.21 & $_{3.01}^{4.73}$ & $^{-6}$
                    & 2.61 & $_{1.39}^{2.63}$ & $^{-5}$
		    & 1.36 & $_{1.13}^{2.13}$ & $^{-6}$  
                    & 4.99 & $_{0.0660}^{\infty}$ & $^{-5}$ \\        
			      
      PG\,0834+500  & 2.91$^{*}$ & $_{1.44}^{3.51}$ & $^{-6}$ 
                    & 6.93 & $_{4.35}^{\infty}$ & $^{-5}$ 
		    & 6.89 & $_{2.84}^{5.71}$ & $^{-6}$
		    & \multicolumn{3}{c}{$\infty$} \\
			      
      PN\,A66\,31   & 2.11 & $_{1.41}^{2.10}$ & $^{-5}$
                    & 4.09 & $_{1.52}^{5.50}$ & $^{-5}$ 
		    & 1.24 & $_{0}^{14.9}$ & $^{-7}$ 
		    & 5.45 & $_{0}^{50.2}$ & $^{-7}$ \\
			      
      HS\,1136+6646 & 1.83$^{*}$ & $_{0.418}^{0.488}$ & $^{-6}$ 
                    & 3.19$^{*}$ & $_{1.24}^{1.29}$ & $^{-5}$ 
		    & 6.67$^{*}$ & $_{1.62}^{6.46}$ & $^{-7}$
                    & \multicolumn{3}{c}{$\infty$} \\           
			      
      Feige\,55     & 2.84$^{*}$ & $_{0.199}^{0.197}$ & $^{-6}$ 
                    & 5.01$^{*}$ & $_{0.311}^{0.542}$ & $^{-6}$
		    & 2.11$^{*}$ & $_{0.103}^{0.205}$ & $^{-6}$
                    & \multicolumn{3}{c}{$\infty$} \\           
			      
      PG\,1210+533  & \multicolumn{3}{c}{}
                    & 4.30 & $_{2.60}^{2.67}$ & $^{-6}$
		    & \multicolumn{3}{c}{} 
                    & 1.50 & $_{0.641}^{1.46}$ & $^{-6}$ \\
		    
      LB\,2         & 2.26 & $_{1.22}^{1.17}$ & $^{-5}$
                    & 2.28 & $_{1.27}^{3.05}$ & $^{-5}$ 
		    & 2.51$^{*}$ & $_{0.864}^{1.40}$ & $^{-6}$
		    & \multicolumn{3}{c}{$\infty$} \\
			      
      HZ\,34        & 1.27 & $_{0.563}^{3.56}$ & $^{-6}$
                    & 4.01 & $_{1.43}^{9.51}$ & $^{-6}$
		    & 2.74 & $_{2.59}^{12.7}$ & $^{-6}$
		    & 3.21 & $_{3.08}^{\infty}$ & $^{-5}$ \\  
			      
      PN\,A66\,39   & 1.66 & $_{1.16}^{2.18}$ & $^{-6}$
                    & 1.66 & $_{0.952}^{21.4}$ & $^{-6}$ 
		    & 5.60 & $_{5.06}^{\infty}$ & $^{-6}$ 
		    & 4.08 & $_{3.87}^{\infty}$ & $^{-5}$ \\
		            			      
      RE\,2013+400  & 1.61 & $_{1.49}^{3.09}$ & $^{-6}$ 
                    & 1.61 & $_{1.52}^{4.01}$ & $^{-6}$
		    & 8.36 & $_{2.03}^{2.62}$ & $^{-7}$ 
		    & 3.01 & $_{1.11}^{6.36}$ & $^{-6}$ \\        		              			      
      PN\,DeHt\,5   & \multicolumn{3}{c}{}
                    & 7.60 & $_{7.20}^{96.6}$ & $^{-7}$ 
		    & \multicolumn{3}{c}{}
                    & 1.08 & $_{0}^{\infty}$ & $^{-5}$ \\
		    
      GD\,561       & 1.16 $^{*}$ & $_{0.409}^{1.96}$ & $^{-5}$
                    & 9.19 $^{*}$ & $_{5.64}^{\infty}$ & $^{-5}$
		    & \multicolumn{3}{c}{$\infty$} 
                    & \multicolumn{3}{c}{$\infty$} \\             
     \hline
     \end{tabular}
\end{table*}

\begin{table*}
    \contcaption{}
    \begin{tabular}{lr@{$^+_-$}l@{$\times$10}lr@{$^+_-$}l@{$\times$10}lr@{$^+_-$}l@{$\times$10}lr@{$^+_-$}l@{$\times$10}l}
      \hline
      Object & \multicolumn{6}{c}{O/H} & \multicolumn{6}{c}{Si/H} \\
      & \multicolumn{3}{c}{Optical} & \multicolumn{3}{c}{\textit{FUSE}} & \multicolumn{3}{c}{Optical} & \multicolumn{3}{c}{\textit{FUSE}} \\
      \hline
      PN\,A66\,7    & \multicolumn{3}{c}{\#} 
		    & \multicolumn{3}{c}{\#}
		    & 7.54$^{*}$ & $_{3.97}^{3.57}$ & $^{-6}$
                    & 2.27$^{*}$ & $_{1.08}^{1.04}$ & $^{-5}$ \\
		 
      HS\,0505+0112 & \multicolumn{3}{c}{\#}
		    & \multicolumn{3}{c}{\#}
		    & 1.67$^{*}$ & $_{0.498}^{0.695}$ & $^{-6}$
                    & 7.71$^{*}$ & $_{2.94}^{1.77}$ & $^{-5}$ \\

      PN\,PuWe\,1   & 9.58$^{*}$ & $_{6.54}^{6.10}$ & $^{-7}$ 
		    & 2.30$^{*}$ & $_{2.06}^{1.19}$ & $^{-6}$
		    & 9.51$^{*}$ & $_{6.55}^{10.1}$ & $^{-6}$ 
                    & 2.99$^{*}$ & $_{2.09}^{1.11}$ & $^{-5}$ \\
			      			      
      RE\,0720-318  & 2.95 & $_{1.38}^{1.12}$ & $^{-6}$ 
		    & 3.04 & $_{0.877}^{3.18}$ & $^{-7}$
		    & 6.28 & $_{1.26}^{2.47}$ & $^{-7}$ 
                    & 6.06 & $_{1.45}^{1.17}$ & $^{-7}$ \\
			      
      Ton\,320      & \multicolumn{3}{c}{$\infty$}  
		    & 3.05$^{*}$ & $_{1.09}^{0.0786}$ & $^{-6}$
		    & 4.33$^{*}$ & $_{3.16}^{4.33}$ & $^{-7}$
                    & 3.73$^{*}$ & $_{2.02}^{5.44}$ & $^{-6}$ \\  
			      
      PG\,0834+500  & \multicolumn{3}{c}{$\infty$} 
		    & 5.67$^{*}$ & $_{2.77}^{4.28}$ & $^{-7}$
		    & 2.81$^{*}$ & $_{0.912}^{0.758}$ & $^{-6}$ 
                    & 3.00$^{*}$ & $_{0.420}^{0.961}$ & $^{-5}$ \\  
			      
      PN\,A66\,31   & 3.11$^{*}$ & $_{0.798}^{2.76}$ & $^{-5}$
		    & 3.04$^{*}$ & $_{0.902}^{0.649}$ & $^{-5}$
		    & 3.53$^{*}$ & $_{1.51}^{2.71}$ & $^{-6}$
                    & 9.42$^{*}$ & $_{2.42}^{4.33}$ & $^{-6}$ \\
			      
      HS\,1136+6646 & 4.76$^{*}$ & $_{0.919}^{1.13}$ & $^{-5}$ 
		    & 9.60$^{*}$ & $_{0.764}^{1.62}$ & $^{-7}$
		    & 4.40$^{*}$ & $_{0}^{2.96}$ & $^{-8}$
                    & 1.32$^{*}$ & $_{0.748}^{1.50}$ & $^{-6}$ \\ 
			      
      Feige\,55     & 1.03$^{*}$ & $_{0.820}^{1.26}$ & $^{-5}$ 
		    & 3.80$^{*}$ & $_{0.337}^{0.370}$ & $^{-7}$
		    & 5.52$^{*}$ & $_{0.732}^{0.820}$ & $^{-7}$ 
                    & 5.15$^{*}$ & $_{1.02}^{1.01}$ & $^{-6}$ \\          
			      
      PG\,1210+533  & \multicolumn{3}{c}{} 
		    & 2.99 & $_{2.00}^{2.40}$ & $^{-6}$
		    & \multicolumn{3}{c}{} 
                    & 1.90 & $_{0.458}^{0.516}$ & $^{-8}$ \\
		    
      LB\,2         & \multicolumn{3}{c}{$\infty$} 
		    & 3.98$^{*}$ & $_{1.11}^{2.97}$ & $^{-7}$
		    & 1.81 & $_{0.587}^{0.730}$ & $^{-6}$ 
                    & 9.84 & $_{3.53}^{6.69}$ & $^{-6}$  \\           
			      
      HZ\,34        & 3.03 $^{*}$ & $_{0.606}^{0.362}$ & $^{-5}$ 
		    & 3.04$^{*}$ & $_{1.19}^{0.209}$ & $^{-7}$
		    & 4.73 & $_{2.41}^{3.76}$ & $^{-6}$
                    & 5.76 & $_{2.74}^{20.9}$ & $^{-6}$ \\ 
                               
      PN\,A66\,39   & \multicolumn{3}{c}{\#}
		    & \multicolumn{3}{c}{\#}
		    & 2.33 & $_{1.24}^{1.81}$ & $^{-6}$
                    & 3.00 & $_{1.27}^{5.11}$ & $^{-6}$ \\ 
			      
      RE\,2013+400  & \multicolumn{3}{c}{\#} 
		    & \multicolumn{3}{c}{\#}
		    & 3.68 & $_{0.845}^{0.963}$ & $^{-7}$ 
                    & 4.71 & $_{1.06}^{1.31}$ & $^{-7}$ \\
			      
      PN\,DeHt\,5   & \multicolumn{3}{c}{}
	            & 7.00 & $_{1.92}^{13.0}$ & $^{-6}$
		    & \multicolumn{3}{c}{}
                    & 8.22 & $_{4.10}^{11.8}$ & $^{-7}$ \\
		    
      GD\,561       & 2.64$^{*}$ & $_{1.28}^{\infty}$ & $^{-4}$ 
		    & \multicolumn{3}{c}{\#}
		    & 5.24$^{*}$ & $_{2.36}^{2.35}$ & $^{-6}$ 
                    & 9.51$^{*}$ & $_{1.34}^{4.11}$ & $^{-6}$ \\
      \hline
     \end{tabular}
\end{table*}

\begin{table*}
    \contcaption{}
    \begin{tabular}{lr@{$^+_-$}l@{$\times$10}lr@{$^+_-$}l@{$\times$10}lr@{$^+_-$}l@{$\times$10}lr@{$^+_-$}l@{$\times$10}l}
      \hline
      Object & \multicolumn{6}{c}{Fe/H} & \multicolumn{6}{c}{Ni/H} \\
      & \multicolumn{3}{c}{Optical} & \multicolumn{3}{c}{\textit{FUSE}} & \multicolumn{3}{c}{Optical} & \multicolumn{3}{c}{\textit{FUSE}} \\
      \hline
      PN\,A66\,7 & 1.24$^{*}$ & $_{0.744}^{1.17}$ & $^{-5}$
		 & 8.22$^{*}$ & $_{3.82}^{7.25}$ & $^{-5}$  
		 & 0 & $_{0}^{1.63}$ & $^{-6}$ 
                 & 0 & $_{0}^{8.94}$ & $^{-6}$ \\
		 
      HS\,0505+0112 & 8.77$^{*}$ & $_{2.93}^{13.1}$ & $^{-5}$
                    & 9.68$^{*}$ & $_{6.19}^{4.06}$ & $^{-5}$          
		    & 1.53 & $_0^{5.30}$ & $^{-6}$ 
                    & 0 & \multicolumn{2}{l}{$_0^\infty$} \\

      PN\,PuWe\,1   & 2.96 & $_0^{11.2}$ & $^{-6}$
                    & 3.40 & $_0^{16.4}$ & $^{-6}$
		    & 1.23 & $_0^{31.4}$ & $^{-6}$ 
                    & 5.65 & $_0^\infty$ & $^{-6}$ \\
			      			      
      RE\,0720-318  & 1.32 & $_{0}^{6.62}$ & $^{-6}$
		    & 2.44 & $_{0}^{13.3}$ & $^{-6}$
		    & 3.45 & $_{0}^{46.1}$ & $^{-7}$
		    & 7.21 & $_{0}^{102.}$ & $^{-7}$ \\
			      
      Ton\,320      & 1.70$^{*}$ & $_{1.16}^{2.37}$ & $^{-5}$
		    & 8.73$^{*}$ & $_{4.64}^{16.3}$ & $^{-5}$
		    & 5.43 & $_{0}^{65.1}$ & $^{-7}$ 
		    & 1.82 & $_{0}^{29.7}$ & $^{-6}$ \\  
			      
      PG\,0834+500  & 1.82$^{*}$ & $_{0.926}^{3.18}$ & $^{-5}$
		    & 4.75$^{*}$ & $_{3.42}^{4.17}$ & $^{-4}$
		    & 8.41 & $_{0}^{42.5}$ & $^{-6}$
		    & 2.21 & $_{0}^{\infty}$ & $^{-5}$ \\  
			      
      PN\,A66\,31   & 9.42$^{*}$ & $_{5.52}^{8.36}$ & $^{-5}$
		    & 3.16$^{*}$ & $_{1.52}^{2.20}$ & $^{-4}$
		    & 2.22 & $_{0}^{39.5}$ & $^{-7}$ 
		    & 7.36 & $_{0}^{234.}$ & $^{-7}$ \\
			      
      HS\,1136+6646 & 5.38$^{*}$ & $_{0}^{11.5}$ & $^{-6}$ 
		    & 0$^{*}$ & $_{0}^{2.31}$ & $^{-6}$ 
		    & 0$^{*}$ & $_{0}^{5.89}$ & $^{-7}$ 
		    & 0$^{*}$ & $_{0}^{2.48}$ & $^{-5}$ \\ 
			      
      Feige\,55     & 1.40$^{*}$ & $_{0.439}^{0.616}$ & $^{-5}$ 
                    & 1.41$^{*}$ & $_{0.358}^{0.480}$ & $^{-5}$
		    & 5.55 & $_{3.73}^{8.54}$ & $^{-6}$
		    & 3.71 & $_{2.21}^{6.22}$ & $^{-6}$ \\          
			      
      PG\,1210+533  & \multicolumn{3}{c}{}  
                    & 4.70$^{*}$ & $_{0}^{60.1}$ & $^{-6}$
		    & \multicolumn{3}{c}{}  
	            & 3.29$^{*}$ & $_{0}^{19.2}$ & $^{-6}$ \\
		    
      LB\,2         & 1.73$^{*}$ & $_{1.10}^{1.92}$ & $^{-5}$
                    & 2.09 & $_{1.06}^{2.42}$ & $^{-5}$ 
		    & 0 & $_{0}^{3.12}$ & $^{-6}$
		    & 0 & $_{0}^{7.37}$ & $^{-6}$ \\           
			      
      HZ\,34        & 2.42 & $_{1.68}^{5.82}$ & $^{-5}$
		    & 7.64 & $_{5.91}^{51.0}$ & $^{-5}$
		    & 0 & $_{0}^{6.63}$ & $^{-6}$ 
		    & 5.12 & $_{0}^{1920.}$ & $^{-8}$ \\ 
                               
      PN\,A66\,39   & 4.89 & $_{0}^{21.4}$ & $^{-6}$
		    & 3.74 & $_{0}^{15.3}$ & $^{-6}$
		    & 4.37 & $_{0}^{25.4}$ & $^{-6}$
		    & 9.93 & $_{0}^{88.7}$ & $^{-6}$ \\ 
			      
      RE\,2013+400  & 0 & $_{0}^{2.83}$ & $^{-5}$
		    & 0 & $_{0}^{2.16}$ & $^{-5}$
		    & 0 & $_{0}^{2.93}$ & $^{-5}$ 
		    & 0 & $_{0}^{3.11}$ & $^{-5}$ \\
			      
      PN\,DeHt\,5   & \multicolumn{3}{c}{}
                    & 3.34 & $_{2.58}^{6.20}$ & $^{-5}$ 
		    & \multicolumn{3}{c}{}
	            & 1.21 & $_{1.19}^{7.27}$ & $^{-5}$ \\
		    
      GD\,561       & 1.55 & $_{0.967}^{1.76}$ & $^{-5}$
		    & 1.45 & $_{0.828}^{0.143}$ & $^{-5}$ 
		    & 1.84 & $_{0}^{11.2}$ & $^{-6}$
		    & 1.78 & $_{0}^{28.1}$ & $^{-6}$ \\
      \hline
     \end{tabular}
\end{table*}

\begin{table*}
  \caption{Summary of which lines were observed to be present when performing the abundance fits.  H$_2$\ denotes where lines were obscured by molecular hydrogen absorption.}
  \label{detections}
  \begin{tabular}{lccccccccccc}
  \hline
  Object & C\,\textsc{iv} & N\,\textsc{iii} & N\,\textsc{iv} & O\,\textsc{iv} & O\,\textsc{vi} & Si\,\textsc{iii} & Si\,\textsc{iv} & Fe\,\textsc{v} & Fe\,\textsc{vi} & FE\,\textsc{vii} & Ni\,\textsc{v} \\
  \hline
 PN\,A66\,7     & $\surd$ & H$_2$ & $\surd$ & $\surd$ & $\surd$ & & $\surd$ & & $\surd$ & $\surd$ & \\
 HS\,0505+0112  & $\surd$ & $\surd$ & $\surd$ & $\surd$ & $\surd$ & & $\surd$ & & & $\surd$ & \\ 
 PN\,PuWe\,1    & $\surd$ & H$_2$ & $\surd$ & $\surd$ & $\surd$ & & $\surd$ & & & & \\
 RE\,0720-318   & $\surd$ & $\surd$ & $\surd$ & & $\surd$ & $\surd$ & $\surd$ & & & & \\
 TON\,320       & $\surd$ & & $\surd$ & $\surd$ & $\surd$ & & $\surd$ & & $\surd$ & $\surd$ & \\
 PG\,0834+500   & $\surd$ & & $\surd$ & $\surd$ & $\surd$ & $\surd$ & $\surd$ & $\surd$ & $\surd$ & $\surd$ & \\
 PN\,A66\,31    & $\surd$ & H$_2$ & $\surd$ & $\surd$ & $\surd$ & & $\surd$ & & $\surd$ & $\surd$ & \\
 HS\,1136+6646  & $\surd$ & & $\surd$ & $\surd$ & $\surd$ & & $\surd$ & & & & \\
 Feige\,55      & $\surd$ & $\surd$ & $\surd$ & $\surd$ & $\surd$ & & $\surd$ & $\surd$ & $\surd$ & & $\surd$ \\
 PG\,1210+533   & $\surd$ & $\surd$ & $\surd$ & $\surd$ & $\surd$ & $\surd$ & $\surd$ & & & \\
 LB\,2          & $\surd$ & & $\surd$ & $\surd$ & $\surd$ & $\surd$ & $\surd$ & & $\surd$ & & \\
 HZ\,34         & $\surd$ & & $\surd$ & $\surd$ & $\surd$ & & $\surd$ & & $\surd$ & $\surd$ & \\
 PN\,A66\,39    & $\surd$ & H$_2$ & $\surd$ & & $\surd$ & & $\surd$ & & & & \\
 RE\,2013+400   & $\surd$ & $\surd$ & $\surd$ & & $\surd$ & $\surd$ & $\surd$ & & & & \\
 PN\,DeHt\,5    & $\surd$ & H$_2$ & $\surd$ & $\surd$ & & $\surd$ & $\surd$ & & & & \\
 GD\,561        & $\surd$ & H$_2$ & $\surd$ & $\surd$ & $\surd$ & & $\surd$ & & $\surd$ & $\surd$ & \\
  \hline
  \end{tabular}
\end{table*} 

\begin{figure}
	\includegraphics[height=\columnwidth,angle=90]{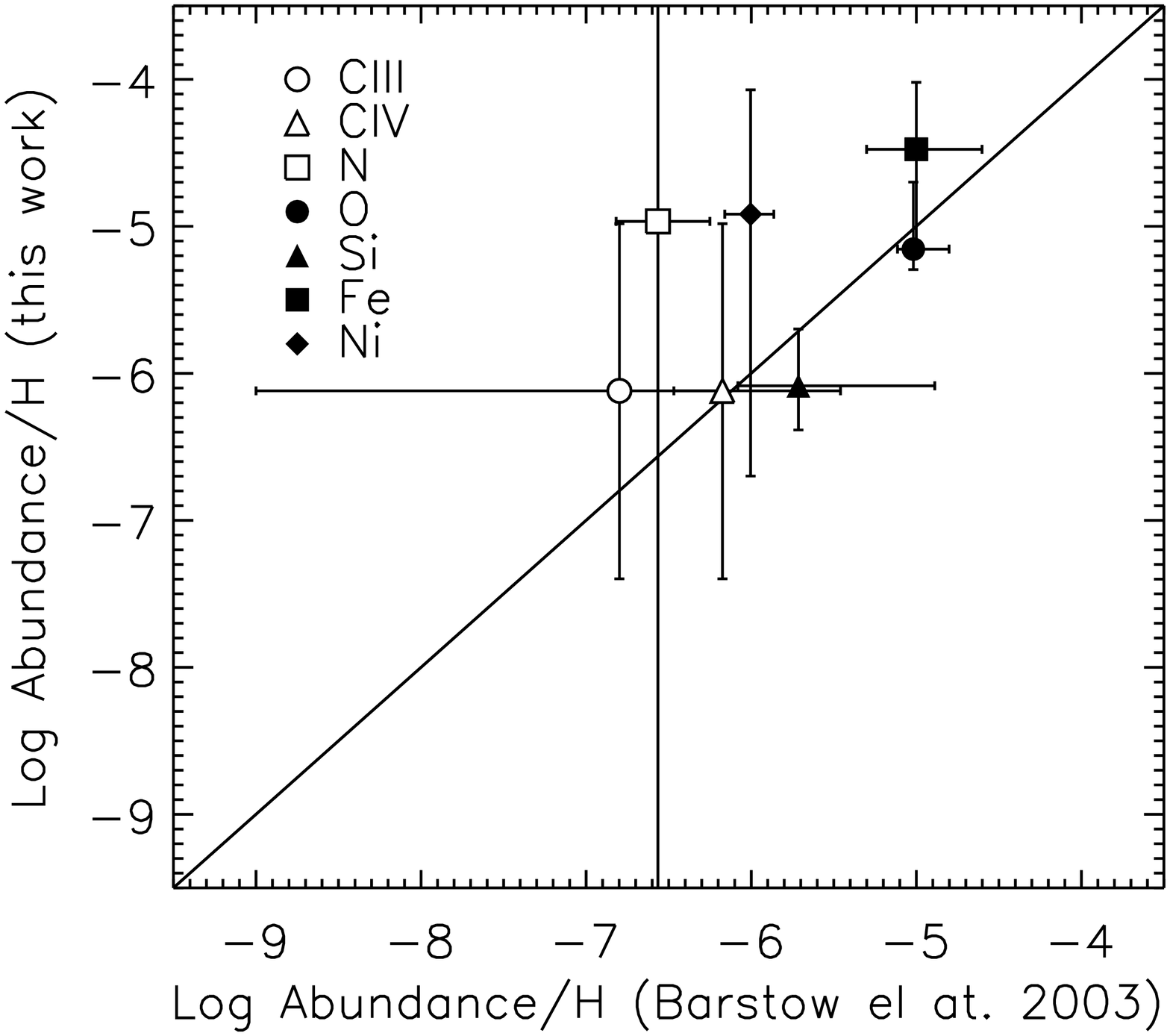} 
	\caption{\label{wd2218} Comparison of abundance measurements for DeHt 5 between \citet{bars03abund}\ and this work.  Note that \citet{bars03abund}\ obtained abundances for C\,{III}\ and C\,{IV}\ separately.  These are both compared to the carbon abundance determined in this work from fits to C\,\textsc{iv}\ lines.}
\end{figure}

The sample of DAOs and the sample of DAs of \citet{bars03abund}\ contain one
object in common: PN\,DeHt\,5.  Although this white dwarf does not have an
observable helium line in its optical spectrum, it is included within the
sample of DAOs because \citet{bars01wd2218}\ detected helium within the
\textit{Hubble Space Telescope}\ (\textit{HST}) \textit{Space Telescope Imaging
Spectrometer}\ (\textit{STIS}) spectrum of the object.  This presents an
opportunity to compare the results of this work with the abundances measured by
\citet{bars03abund}\ to confirm the consistency of the measurements made in the
different wavebands; this comparison is shown in Figure \ref{wd2218}, and
demonstrates that agreement between the sets of results is within errors. 
However, uncertainties in the measurements are large, particularly for nitrogen
measured in this work.

\begin{figure}
	\includegraphics[height=\columnwidth,angle=90]{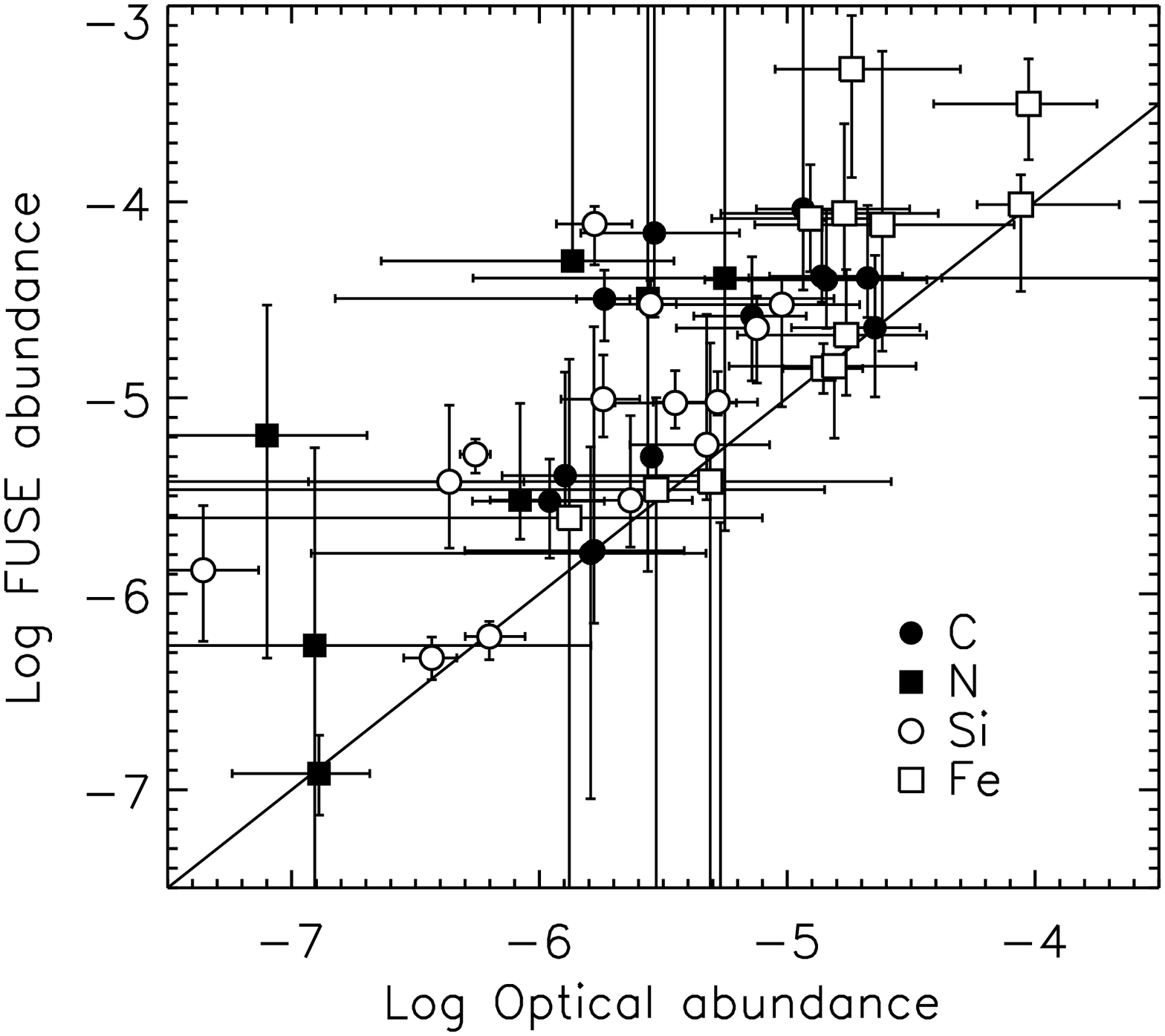} 
	\caption{\label{comparison} Comparison of the abundances of carbon, nitrogen, silicon and iron measured when the temperature, gravity and helium abundance derived from \textit{FUSE}\ and optical data were used.  Also shown is a line marking where the abundances are equal.}
\end{figure}

The abundances measured with the models that use the \textit{FUSE}\ and optical measurements of temperature, gravity and helium abundance can also be compared from the results of this work.  Changing the parameters of the model will affect its temperature structure and ionisation balance, and therefore the abundances measured may also change.  The carbon, nitrogen, silicon and iron abundances measured using the \textit{FUSE}\ derived parameters are compared to those measured using the optically derived parameters in Figure \ref{comparison}.  Oxygen abundances are not shown due to the poor quality of the fits, while many of the nickel abundances are poorly constrained and hence are also not shown.  The abundances that were measured with models that used the \textit{FUSE}\ parameters are generally higher than those that used the optical parameters by a factor of between 1 and 10.  In some cases the factor is higher, for example it is 30 for the silicon measurements for HS\,1136+6646.  However, a satisfactory fit to those lines was not achieved ($\chi^2_{red} > 2$), hence the abundances may be unreliable.  In contrast, the fits to the nitrogen lines in the spectrum of A\,7 were satisfactory, yet the abundance measure that used the \textit{FUSE}\ parameters was higher than the measure with optical parameters by a factor of 81, although the error bars are large.  Overall, however, the abundance measures that utilise the different parameters appear to correlate with each other.

In the following we describe the abundance measurements for each element in turn, compare them to the DA measurements made by \citet{bars03abund}\ and discuss them with reference to the radiative levitation predictions of \citet{chay95}, and the mass loss calculations of \citet{ungl00}.

\subsection{Carbon abundances}

\begin{figure}
	\includegraphics[height=\columnwidth,angle=90]{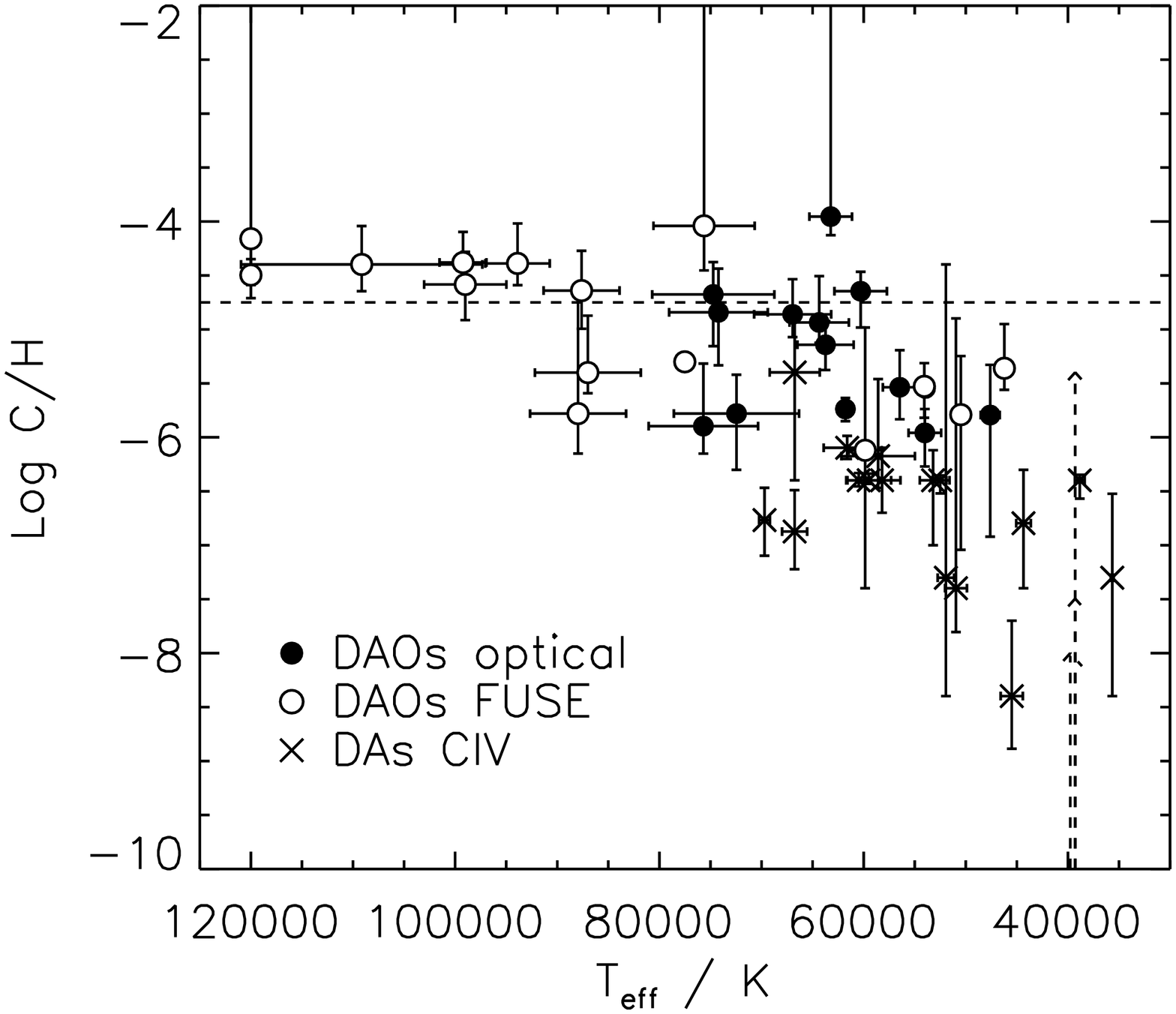}
	\caption{\label{c_teff} Carbon abundances measured for the DAOs against \teff, when the temperature, gravity and helium abundance were fixed according to the values measured by \citet{good04}\ from fits to hydrogen Balmer lines (filled circles) and hydrogen Lyman lines (open circles).  Also shown for comparison are the results of \citet{bars03abund}\ for DA white dwarfs (crosses).  The dashed arrows mark the 3$\sigma$\ upper abundance limit for the DAs, where carbon was not detected.  The approximate position of the abundances predicted by \citet{chay95}\ for a DA white dwarf with a \logg\ of 7.0 is shown as a dashed line.}
\end{figure}

Figure \ref{c_teff}\ shows the measured carbon abundances against \teff from fits to the carbon absorption features in the \textit{FUSE}\ data. Abundances obtained when the model \teff, \logg\ and helium abundance were set to the values found by \citet{good04}\ from fits to both optical and \textit{FUSE}\ spectra are shown.  In the following, these will be referred to as `optical abundances' and `\textit{FUSE}\ abundances' respectively.  Shown for comparison are the measurements made by \citet{bars03abund}\ for DAs.  For clarity, only the result of their fits to the C\,\textsc{iv}\ lines are plotted.  The plot shows that the carbon abundances of the DAOs fall within the range 10$^{-6}$\ to 10$^{-4}$\ that of hydrogen, in general higher than those for the DAs.  The only abundance below 10$^{-6}$\ belongs to PN\,DeHt\,5, which, apart from the discovery of helium in its \textit{HST}\ spectrum by \citet{bars01wd2218}, would not be classed as a DAO in this work.  The other DAO white dwarfs whose temperatures are similar to those of the DAs have marginally higher abundances than the latter.  These DAOs comprise the white dwarf plus main sequence star binaries, the unusual white dwarf PG\,1210+533, and optical abundances of stars whose \teff, as measured from \textit{FUSE}\ data, were extreme \citep[see][]{good04}.  However, the highest abundance measured was for HS\,0505+0112, which is also one of the extreme \textit{FUSE}\ \teff\ objects, for which the \textit{FUSE}\ abundance and the upper limit of the optical abundance exceeded the upper limit of the model grid.

The predictions of \citet{chay95}\ suggest that above $\sim$40\,000 K, carbon abundances should have little dependence on temperature, but should increase with decreasing gravity.  Taking the optical abundances together with the DA abundances, it might be argued that they form a trend of increasing temperature with gravity, extending to $\sim$80\,000 K.  Since the measurements of \teff\ from \textit{FUSE}\ were higher than those from optical data, the \textit{FUSE}\ abundances are more spread out towards the higher temperatures than the optical abundances.  The \textit{FUSE}\ abundances do not show any trend with temperature, although the lower temperature objects and three others do have slightly lower abundances than the remainder; if the latter objects (Feige\,55, HZ\,34 and PN\,A\,39) had high \logg, this might explain their low carbon abundances, but this does not seem to be the case.  Overall, no carbon abundance was greater than $\sim$10$^{-4}$ that of hydrogen, which is quite close to the maximum abundance predicted by \citet{chay95}, illustrated in Figure \ref{c_teff}\ by the dashed line.  However the carbon abundances decrease below $\sim$80\,000 K, contrary to their predictions. \citet{ungl00}\ predict a surface abundance of carbon of between 10$^{-3}$\ and 10$^{-4}$\ times the number of heavy particles in their simulation.  Their calculations end at their wind limit, when the wind is expected to cease, at approximately 90\,000 K.  After this point, \citet{ungl00}\ expect the abundances to fall to the values expected when there is equilibrium between gravitational settling and radiative acceleration.  When using an alternative prescription for mass loss, which does not have a wind limit, the drop off occurs at lower temperature, at approximately 80\,000 K.  This is similar to what is observed, although the abundance measurements are a factor $\sim$10 smaller than predicted.  

\begin{figure}
	\includegraphics[height=\columnwidth,angle=90]{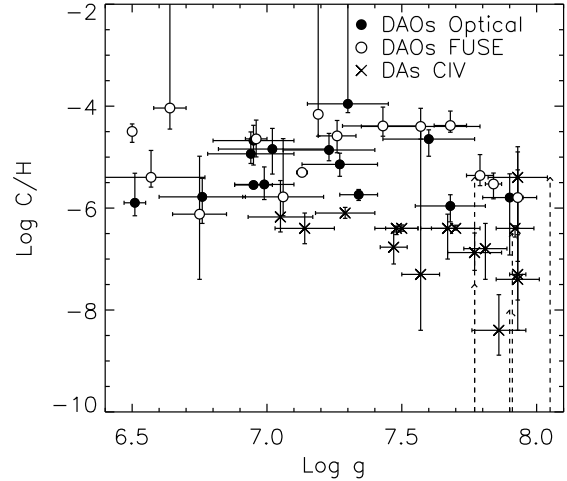}
	\caption{\label{c_logg} Carbon abundances measured for the DAOs against \logg, when the temperature, gravity and helium abundance were fixed according to the values measured by \citet{good04}\ from fits to hydrogen Balmer lines (filled circles) and hydrogen Lyman lines (open circles).  Also shown for comparison are the results of \citet{bars03abund}\ for DA white dwarfs (crosses).  The dashed arrows mark the 3$\sigma$\ upper abundance limit for the DAs, where carbon was not detected.}
\end{figure}

\begin{figure}	 
	\includegraphics[height=\columnwidth,angle=90]{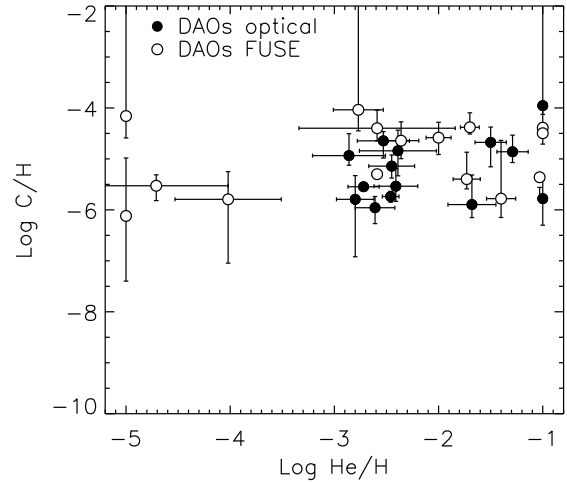}    
	\caption{\label{c_he} Carbon abundances measured for the DAOs against log~$\frac{He}{H}$, when the temperature, gravity and helium abundance were fixed according to the values measured by \citet{good04}\ from fits to hydrogen Balmer lines (filled circles) and hydrogen Lyman lines (open circles).}
\end{figure}

The dependence of carbon abundance on gravity is shown in Figure \ref{c_logg}.  \citet{good04}\ found no systematic differences between \logg\ measured from optical and \textit{FUSE}\ data, hence there is no separation between the optical and \textit{FUSE}\ abundances, as seen in Figure \ref{c_teff}.  No trend between \logg\ and carbon abundance is evident, in contrast to what might be expected from the results of \citet{chay95}.  Figure \ref{c_he}\ shows the relationship between carbon and helium abundance.  As with Figure \ref{c_logg}, no trend is evident.  Four of the points are separated from the rest because of their low helium abundance measured from optical data; this is discussed in \citet{good04}\ and does not appear to be reflected in the carbon abundance.

\subsection{Nitrogen abundances}

\begin{figure}
	\includegraphics[height=\columnwidth,angle=90]{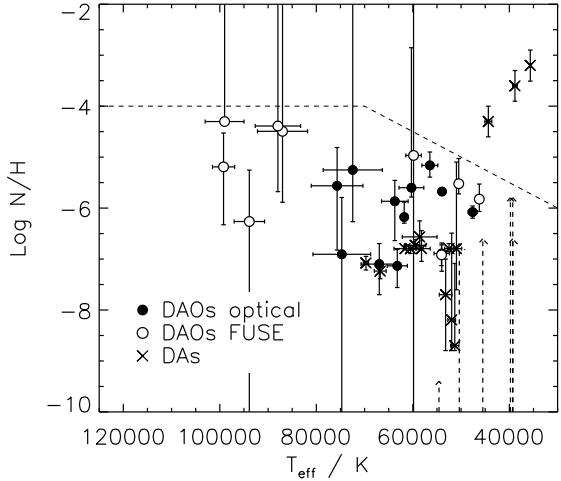}    
	\caption{\label{n_teff} As Figure \ref{c_teff}, but showing nitrogen abundances.}
\end{figure}

Figure \ref{n_teff}\ shows the nitrogen abundance measurements, plotted against effective temperature.  A number of the abundances are relatively poorly constrained, and exceed the upper bound of the model grid.  In contrast, the lower 3$\sigma$\ error bounds for PN\,A\,31\ and PN\,DeHt\,5 both reach below the lowest abundance in the model grid, with, in the latter case, the error bounds extending across the whole range of the model grid.  There is no obvious trend between \teff\ and either the optical or \textit{FUSE}\ abundances, in contrast to the predictions of \citet{chay95}\ and \citet{ungl00}, although the abundances are close to those predicted from radiative levitation theory.  None of the lower temperature DAOs have the comparatively high abundances found for three of the DAs of \citet{bars03abund}.  As with carbon, the DAO nitrogen abundances are often higher than those of the DAs.  However, the lower temperature DAOs do not, in general, have nitrogen abundances different from the higher temperature objects, despite the possible differences in their evolutionary paths.  RE\,0720-318 has a nitrogen abundance similar to the DAs, and less than most of the DAOs, according to the abundances determinations using both the optical and \textit{FUSE}\ parameters.  Three other objects (PN\,A\,7, HS\,0505+0112 and PN\,A\,31) have comparatively low abundances when the optically determined parameters were used.  However, this was not the case when the \textit{FUSE} parameters were used.

\begin{figure}
	\includegraphics[height=\columnwidth,angle=90]{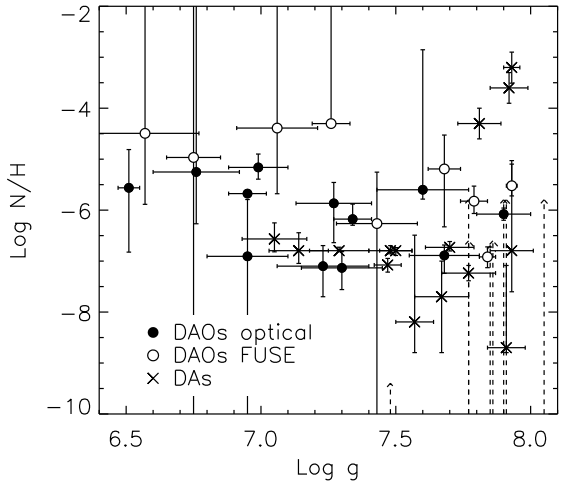}
	\caption{\label{n_logg} As Figure \ref{c_logg}, but showing nitrogen abundances.}    
\end{figure}

\begin{figure}
	\includegraphics[height=\columnwidth,angle=90]{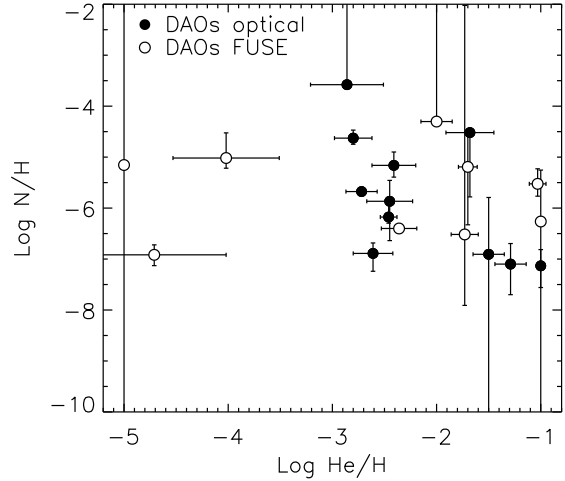}    
	\caption{\label{n_he} As Figure \ref{c_he}, but showing nitrogen abundances.}
\end{figure}

In Figures \ref{n_logg}\ and \ref{n_he}, nitrogen abundances are plotted against \logg\ and log~$\frac{He}{H}$.  As for carbon, no trend with gravity or helium abundance is evident.

\subsection{Oxygen abundances}

\begin{figure}
	\includegraphics[height=\columnwidth,angle=90]{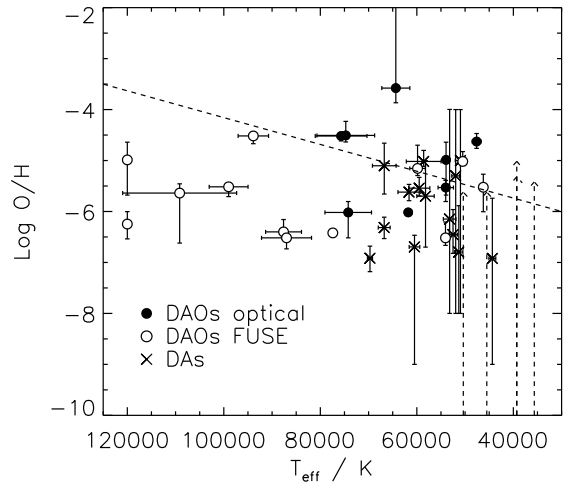}
	\caption{\label{o_teff} As Figure \ref{c_teff}, but showing oxygen abundances.}
\end{figure}

All the DAOs were found to have evidence of oxygen absorption in their spectra.  This absorption is believed to originate in the white dwarf as the radial velocity of the lines are not consistent with those of interstellar absorption features.  The DAO oxygen abundances, shown in Figure \ref{o_teff} against \teff, are very similar to those of the DAs and show no obvious dependence on temperature.  Again, the lower temperature DAOs do not appear to have abundances different from the higher temperature objects.  However, the reproduction of the observed oxygen lines by the models was very poor, with the problem affecting both high and low temperature DAOs.  For only three objects, RE\,0720-318, PG\,1210+533 and PN\,DeHt\,5, was a fit with an acceptable value of $\chi^2_{red}$\ ($<$~2) achieved.  In general the lines predicted by the models were too weak compared to the observed lines.  In four cases an abundance was not even recorded for either optical or \textit{FUSE}\ abundances, and in one case (GD\,561) for the \textit{FUSE}\ abundance alone, due to the disagreement between the model and data (see, for example, Figure \ref{re2013fuse_o}).  The poor quality of the fits means that the results are probably not reliable, and we do not show plots of abundance against gravity or helium abundance.

\subsection{Silicon abundances}

\begin{figure}
	\includegraphics[height=\columnwidth,angle=90]{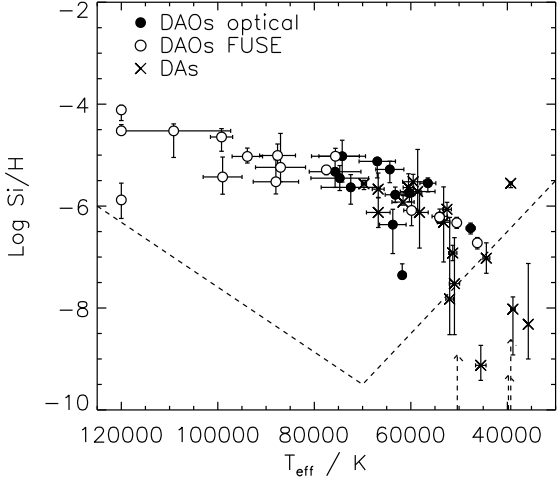}
	\caption{\label{si_teff} As Figure \ref{c_teff}, but showing silicon abundances.}
\end{figure}

\begin{figure}	    
	\includegraphics[height=\columnwidth,angle=90]{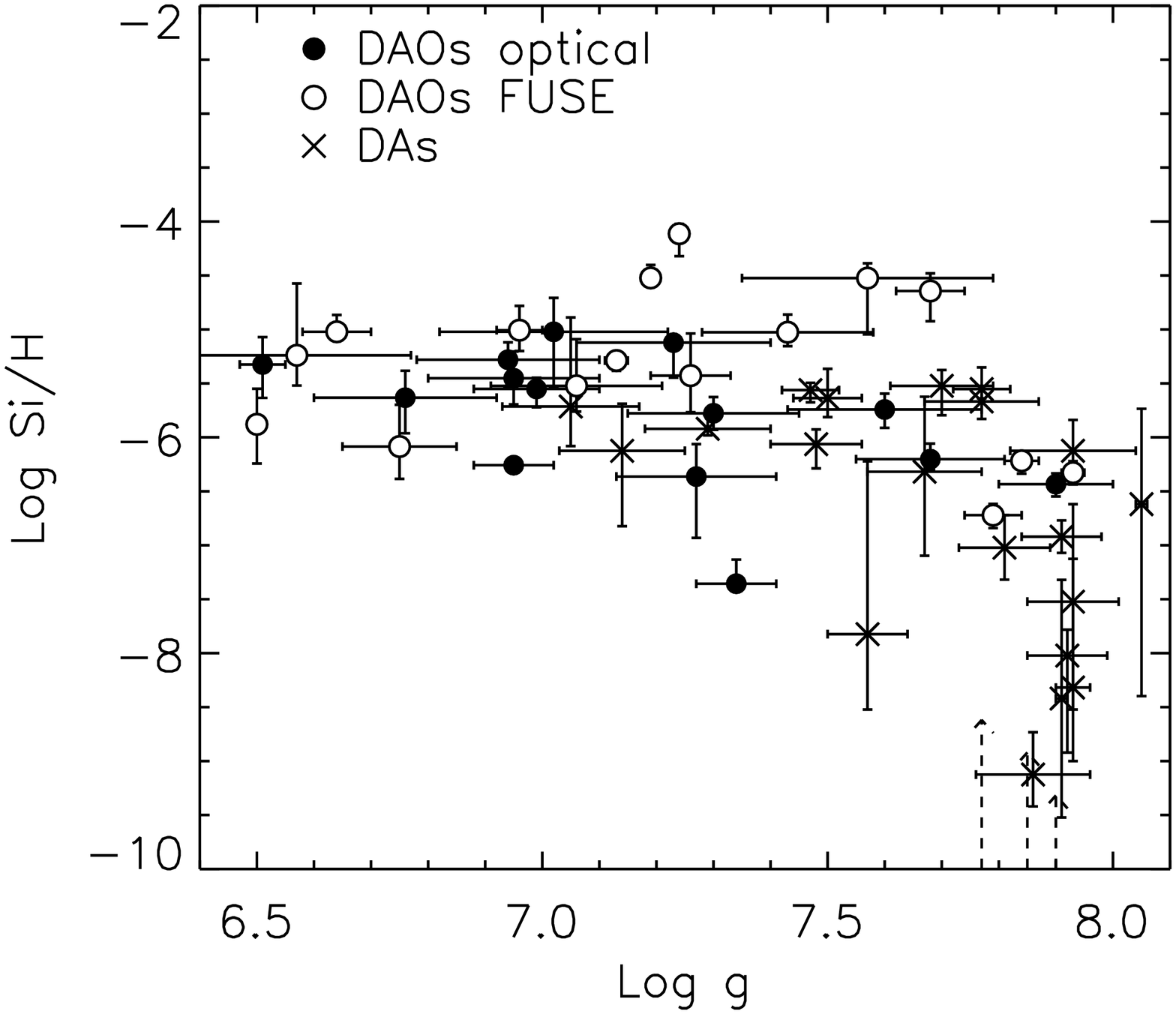}    
	\caption{\label{si_logg} As Figure \ref{c_logg}, but showing silicon abundances.}    
\end{figure}

\begin{figure}
	\includegraphics[height=\columnwidth,angle=90]{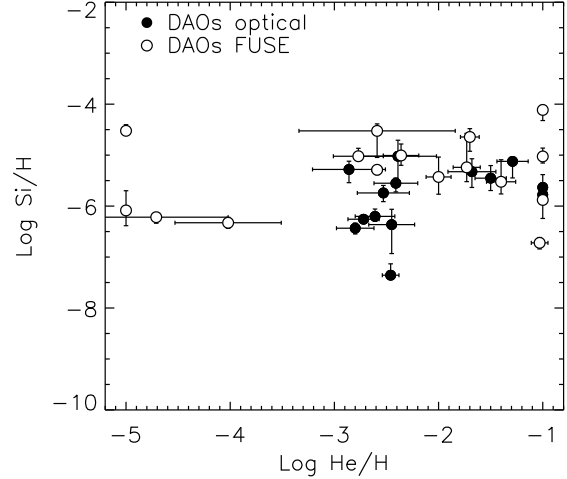}    
	\caption{\label{si_he} As Figure \ref{c_he}, but showing silicon abundances.}
\end{figure}

Figure \ref{si_teff}\ shows the measured silicon abundances against \teff.  The optical abundances, when considered along with the DA abundances, appear to show an increase with temperature.  The \textit{FUSE}\ abundances also show a slight increase with temperature, with the maximum abundance observed $\sim$10$^{-4}$\ that of hydrogen.  The calculations of \citet{chay95}\ predict a minimum at 70\,000 K, which is not observed here, although \citet{bars03abund}\ did note an apparent minimum at $\sim$40\,000 K in the DA abundances.  However, as with the oxygen measurements, the fits were often poor and in approximately half the cases $\chi^2_{red} < $~2 was not achieved.  The abundance measurements for HS\,1136+6646 are lower than for the other DAOs.  This is one of the objects for which satisfactory fits were not achieved; it is also one of the DAOs with extreme \teff, as measured from \textit{FUSE}\ data.  In Figure \ref{si_logg}, we plot the silicon abundances against \logg.  Similarly, Figure \ref{si_he}\ shows the relationship between heavy element abundances and helium abundances.  In neither are any strong trends evident.

\subsection{Iron abundances}

\begin{figure}
	\includegraphics[height=\columnwidth,angle=90]{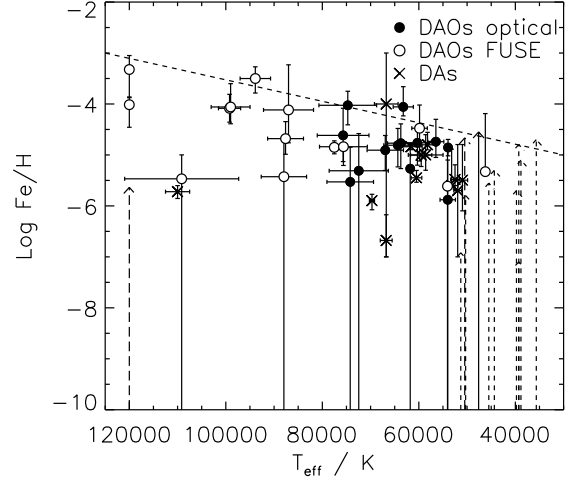}
    	\caption{\label{fe_teff} As Figure \ref{c_teff}, but showing iron abundances.  Solid and long dashed arrows mark the 3$\sigma$\ upper limits for the optical and \textit{FUSE}\ abundances, respectively, where iron was not detected.}
\end{figure}

\begin{figure}
	\includegraphics[height=\columnwidth,angle=90]{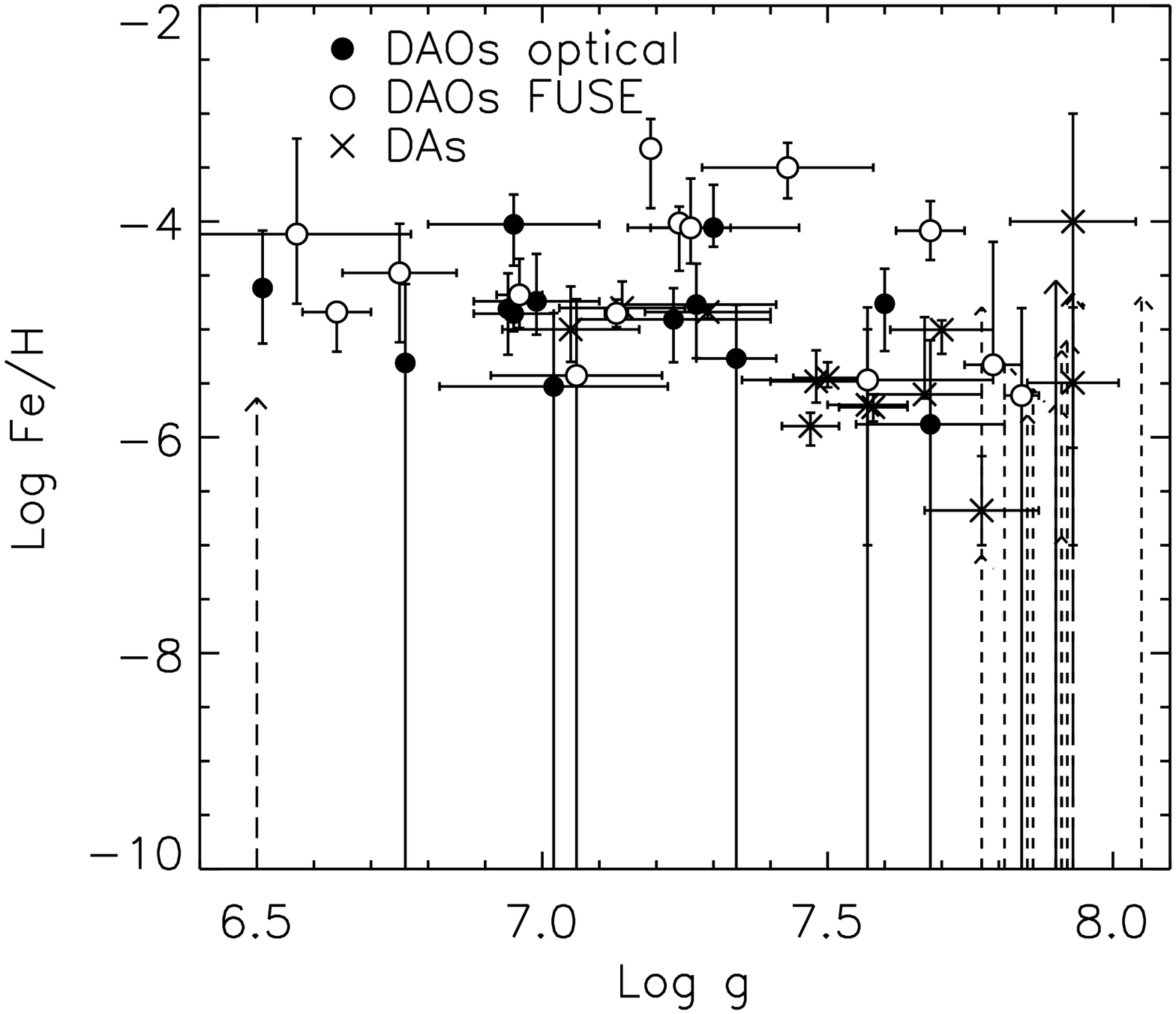}
	\caption{\label{fe_logg} As Figure \ref{c_logg}, but showing iron abundances.  Solid and long dashed arrows mark the 3$\sigma$\ upper limits for the optical and \textit{FUSE}\ abundances, respectively, where iron was not detected.}    
\end{figure}

\begin{figure}    
	\includegraphics[height=\columnwidth,angle=90]{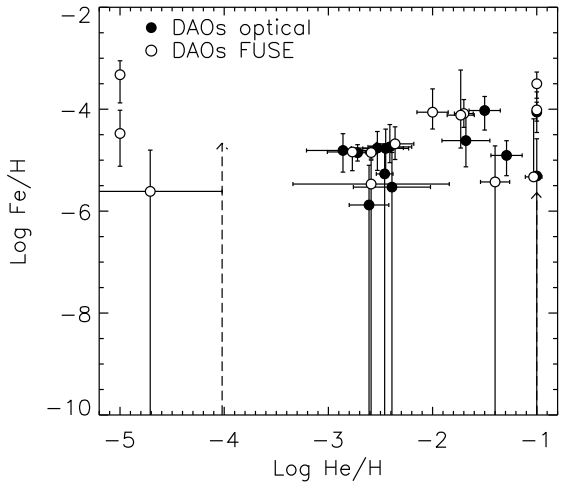}    
	\caption{\label{fe_he} As Figure \ref{c_he}, but showing iron abundances.  Solid and long dashed arrows mark the 3$\sigma$\ upper limits for the optical and \textit{FUSE}\ abundances, respectively, where iron was not detected.}
\end{figure}

Fits to the iron lines were unsatifactory for, again, approximately half the objects.  Apart from three, these were the same objects for which the fits to the silicon lines were also poor.  Included in these are the three objects with extreme \textit{FUSE}-measured \teff, and also PG\,1210+533, which has a temperature and gravity that would be considered normal for a DA white dwarf.  Iron was not predicted in the spectrum of RE\,2013+400, nor that of HS\,1136+6646 when the \textit{FUSE}-derived values of \teff, \logg\ and helium abundance were used.  The lower error boundary reached beyond the lowest abundance in the model grid in eleven cases.  Figure \ref{fe_teff}\ illustrates the relationship between iron abundance and \teff.  This demonstrates that the optical and \textit{FUSE}\ abundances measured for iron are similar to those measured in the DAs, where iron was detected at all.  There is no strong indication of an increase in the abundances with temperature as might be expected from radiative levitation theory.  However, the measured abundances are quite close to the predictions.  No strong differences between the iron abundances of lower and higher temperature DAOs is evident.  Similarly, Figure \ref{fe_logg}\ shows no increase in abundance with decreasing gravity as might also be expected.  Finally, there are also no trends evident in the plot of iron abundances against helium abundances (Figure \ref{fe_he}). 

\subsection{Nickel abundances}

The nickel lines in the \textit{FUSE}\ wavelength range that were used for abundance measurements were weak, hence the results were poorly constrained.  There were a number of non-detections, and the lower bounds for all but three of the measurements were below the range of the model grid.  In addition, three had upper bounds above the highest abundance in the grid and thus their error bars span the entire range of the models.  In no cases was the measured abundance above $\log \frac{Ni}{H} > -4$, and no trends are evident.  

\subsection{Iron/nickel ratio}

\citet{bars03abund}\ found the ratio between the iron and nickel abundances for the DAs to be approximately constant, at the solar value of $\sim$20, rather than the prediction of \citet{chay94}\ of close to unity.  For the DAOs, approximately half have Fe/Ni ratio between 1 and 20, and all but two of the remainder have ratios greater than 20.  However these ratios are very poorly constrained by the fits to \textit{FUSE}\ data, thus preventing definite conclusions about the iron/nickel ratio in DAO white dwarfs.

\section{Discussion}
\label{discussion}

The investigation of heavy element abundances in DAO white dwarfs has provided an opportunity to investigate the ability of our homogeneous models to reproduce the observed lines, when temperatures derived from optical and \textit{FUSE}\ data were used.  These are observed to differ greatly in some cases; Table \ref{abunddisagreelist}\ lists those objects for which temperature disagreements were noted by \citet{good04}.  

\begin{table}
  \begin{center}
    \caption{A list of objects whose optical and \textit{FUSE}\ temperatures
    agree and those that disagree, separated into those where an
    empirical relationship between the two can be found, and those
    with \textit{FUSE} temperatures greater than 120\,000\,K.}
    \label{abunddisagreelist}
    \begin{tabular}{lll}
      \hline
      \multicolumn{1}{c}{Agree} & \multicolumn{2}{c}{Disagree} \\
      & \multicolumn{2}{c}{FUSE \teff} \\
       & $<$120,000\,K & $>$120,000\,K \\
      \hline
      RE\,0720-318 & PN\,A66\,7  & HS\,0505+0112 \\
      PG\,1210+533 & PN\,PuWe\,1 & PG\,0834+500 \\
      RE\,2013+400 & Ton\,320    & HS\,1136+6646 \\
      PN\,DeHt\,5  & PN\,A66\,31 \\
                   & Feige\,55 \\
                   & LB\,2 \\
                   & HZ\,34 \\
                   & PN\,A66\,39 \\
                   & GD\,561 \\
      \hline
    \end{tabular}
  \end{center}
\end{table}

\begin{figure*} 
	\includegraphics[]{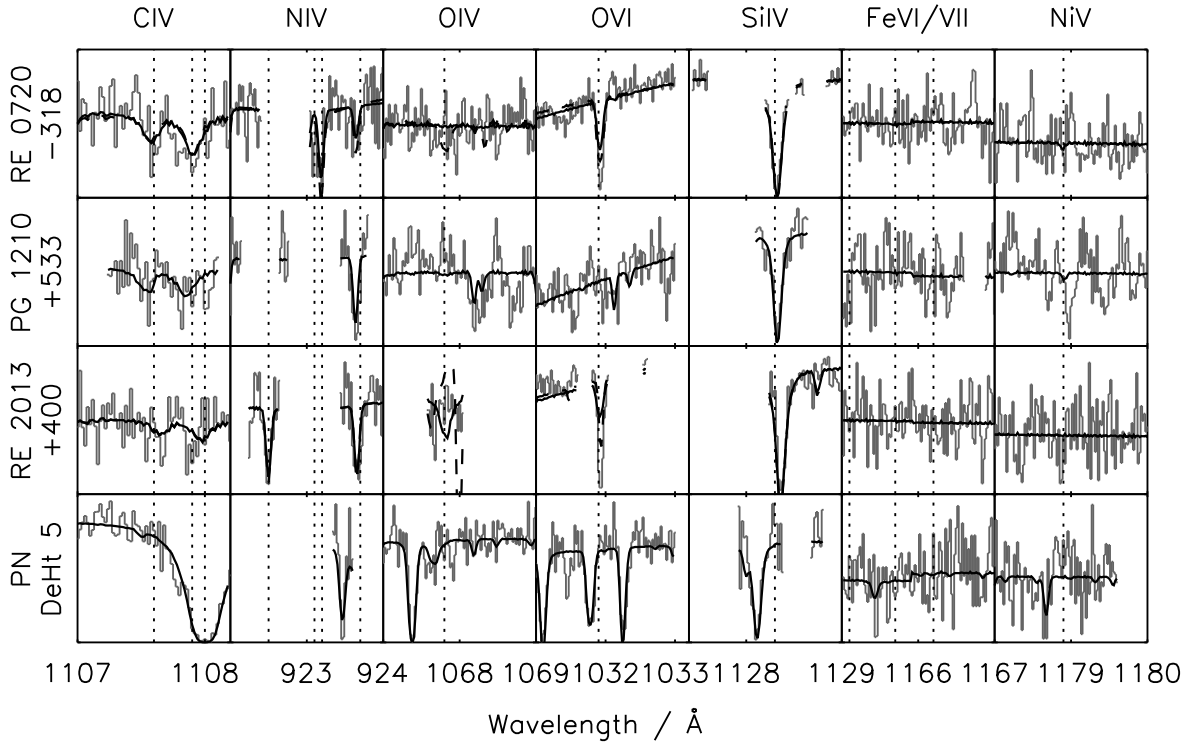} 
	\caption{\label{low_temp_diff_lines} Comparisons of the strengths of lines in the data (grey histograms) to those predicted by the models when the optical (dashed lines) and \textit{FUSE}\ (solid lines) parameters were used, for those objects where the optical and \textit{FUSE}\ temperatures agree.  The vertical axis is scaled individually for each panel to encompass the range of fluxes in that region of the spectrum.  Vertical dashed lines indicate the laboratory wavelength of the lines predicted by the models.  Additional lines seen in some panels (for example PN DeHt 5 C\,\textsc{iv}\ and O\textsc{vi}) are due to molecular hydrogen absorption.  Regions of the spectra containing other photospheric lines, unexplained emission, or where the molecular hydrogen model did not adequately reproduce the shape of the lines, were removed to avoid them influencing the fits.}
\end{figure*}

\begin{figure*} 
	\includegraphics[]{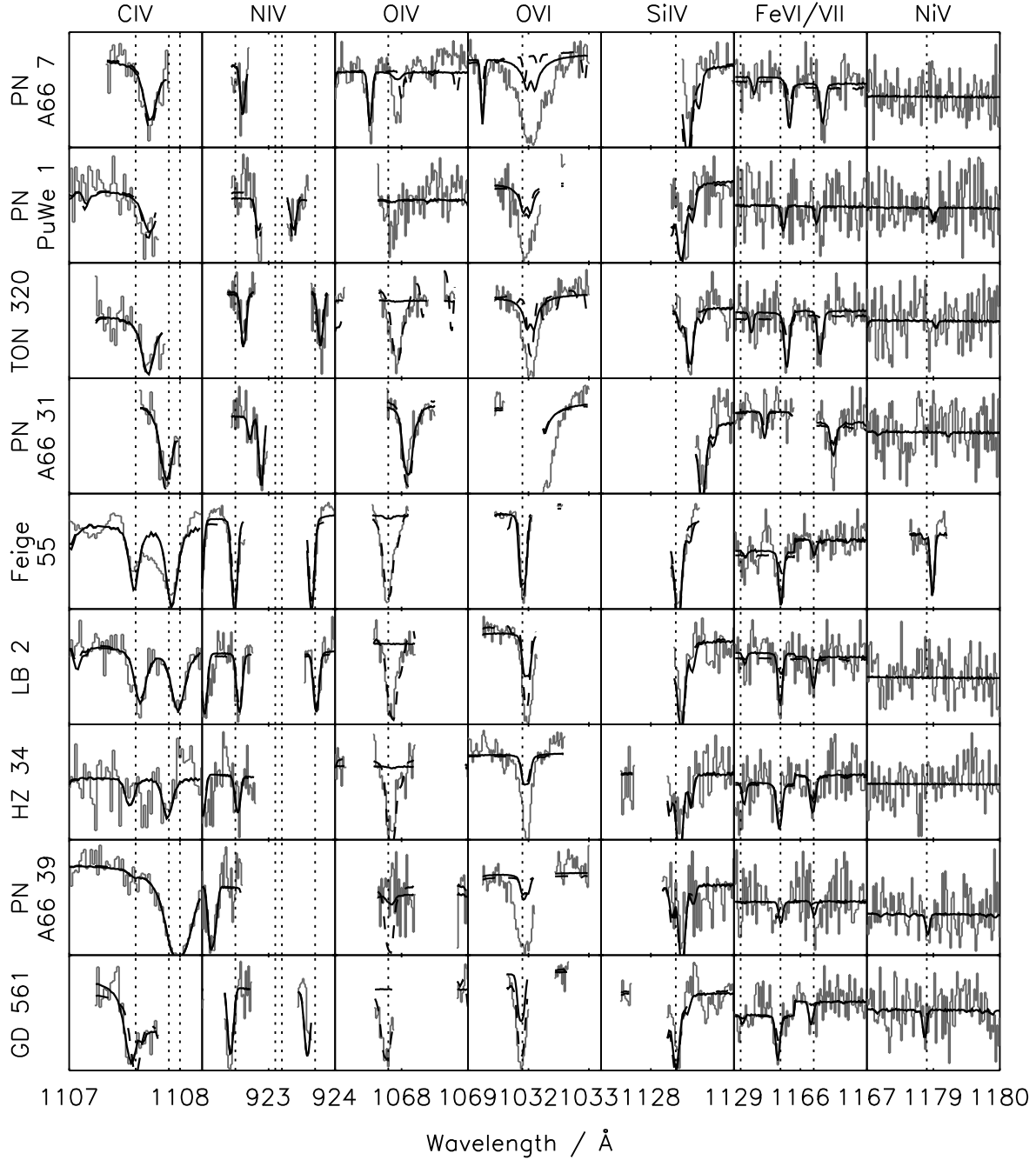} 
	\caption{\label{high_temp_diff_lines} As Figure \ref{low_temp_diff_lines}\ for those objects where the optical and \textit{FUSE}\ temperatures do not agree.}
\end{figure*}

\begin{figure} 
	\includegraphics[]{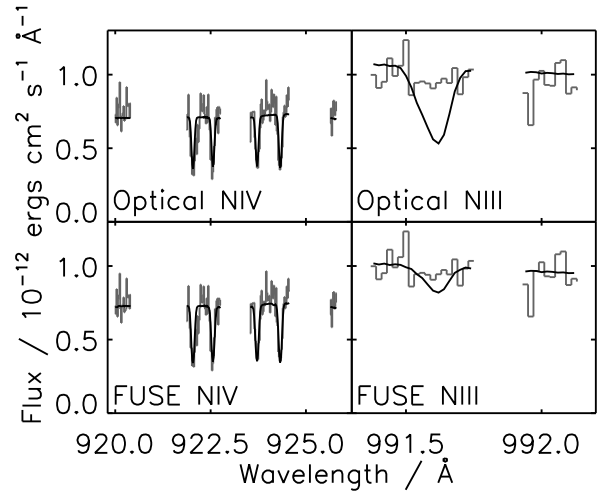} 
	\caption{\label{lb2_line_strengths} Comparisons of line strengths predicted by the models (black line) to the data (grey histogram) for N\,\textsc{iii}\ and N\,\textsc{iv}\ lines in the spectrum of LB\,2, when \teff, \logg\ and log $\frac{He}{H}$\ determined from Balmer and Lyman line analyses are used.}
\end{figure}

In \S\ref{results}, it was found that when the \textit{FUSE}\ values of temperature, gravity and helium abundance were used, the abundances tended to be between a factor 1 and 10 greater than those where the optical values are utilised.  Inspection of Table \ref{abundanceresults}\ shows that abundances of heavy elements, or their upper limits, exceeded the quantities within the model grid more often in the \textit{FUSE}\ results, for example for the carbon measurement of HS\,0505+0112, which is one of the extreme temperature objects.  Satisfactory fits ($\chi^2_{red} <$\ 2) were achieved in approximately equal number when the two sets of parameters were used.  Figure \ref{low_temp_diff_lines}\ shows examples of fits to the spectra of objects where the optical and \textit{FUSE}\ temperatures agree closely.  As might be expected, little difference is seen between the ability of the models that use the optical and \textit{FUSE}\ parameters to reproduce the data.  For the objects where there are differences between the optical and \textit{FUSE}\ temperatures (Figure \ref{high_temp_diff_lines}), large differences are seen between the reproduction of the oxygen lines.  The models that use the comparatively high \textit{FUSE}\ temperatures do not have strong O\,\textsc{iv}\ lines.  However, the models that use the optical temperatures are able to reproduce the lines.  The O\,\textsc{vi}\ lines are not well reproduced by the models, except in a few cases, for example GD\,561.  In contrast, the iron lines tend to be better reproduced by the models that use the \textit{FUSE}\ data, as the lines predicted by the model with optical parameters are too weak.  In addition, Figure \ref{lb2_line_strengths}\ compares the strengths of N\,\textsc{iii}\ and N\,\textsc{iv}\ lines in the spectrum of LB\,2 when the optical and \textit{FUSE}\ parameters were adopted.  Although in neither case was the N\,\textsc{iii}\ line well reproduced, the model that uses the \textit{FUSE}\ parameters is closer to what is observed.  

\begin{figure*} 
	\includegraphics[]{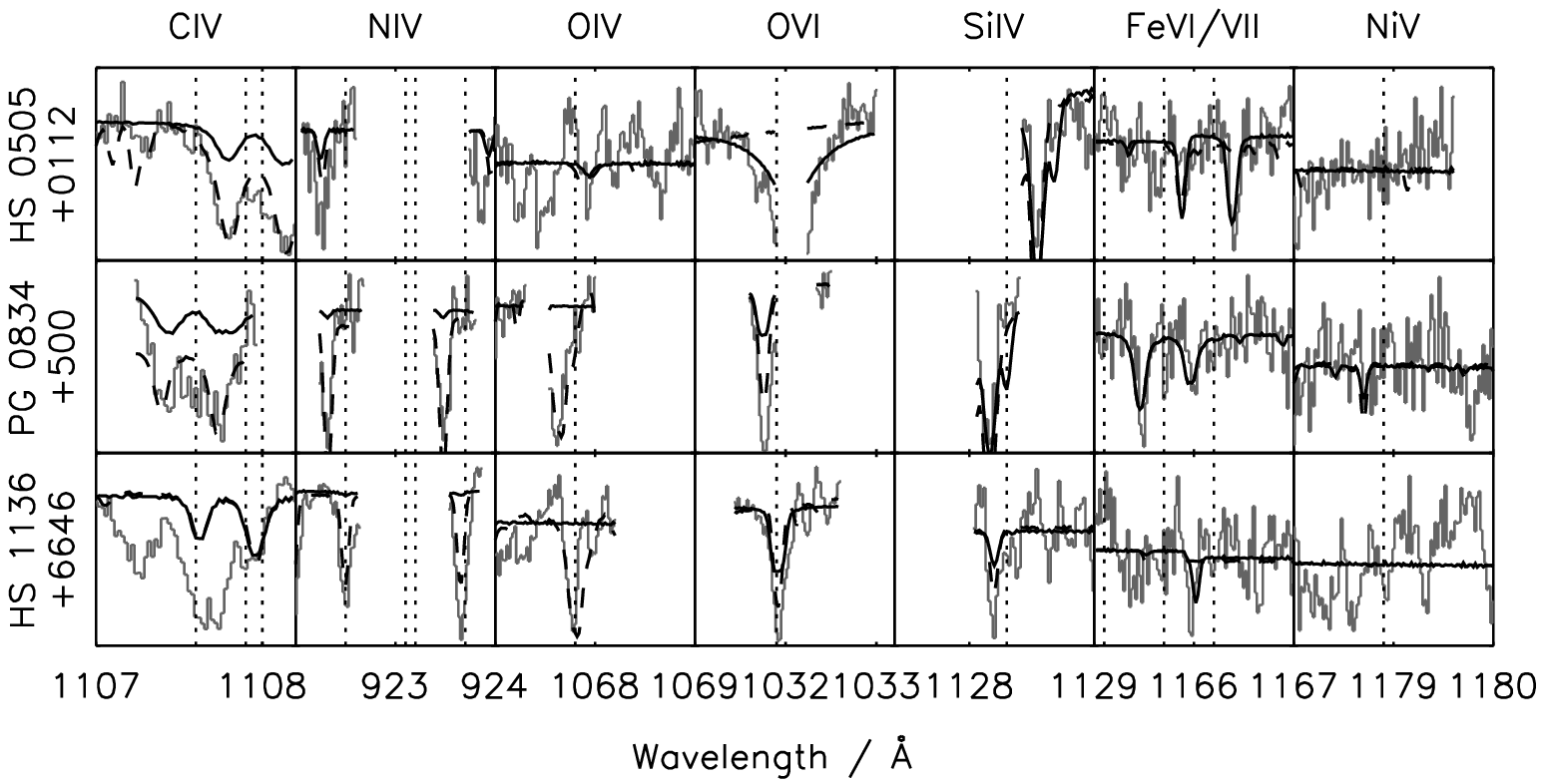} 
	\caption{\label{extreme_temp_diff_lines} As Figure \ref{low_temp_diff_lines}\ for those objects where the \textit{FUSE}\ temperature exceeded 120\,000 K.}
\end{figure*}

\begin{figure}
	\includegraphics[]{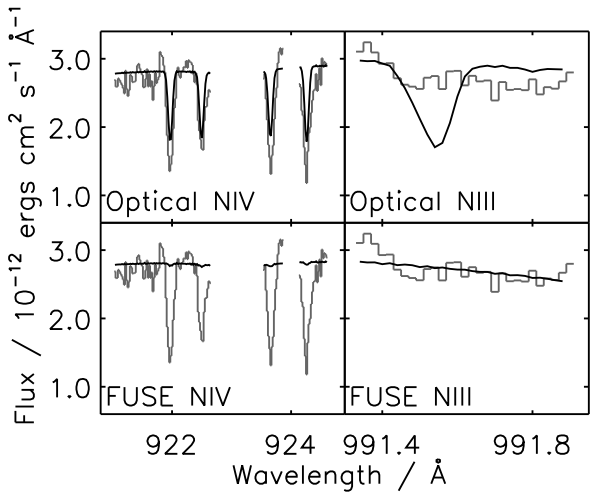}
	\caption{\label{hs1136_line_strengths} Comparisons of line strengths predicted by the models (black line) to the data (grey histogram) for N\,\textsc{iii}\ and N\,\textsc{iv}\ lines in the spectrum of HS\,1136+6646, when \teff, \logg\ and log $\frac{He}{H}$\ determined from Balmer and Lyman line analyses are used.}
\end{figure}

Figure \ref{extreme_temp_diff_lines}\ show example fits for the objects where the \textit{FUSE}\ measure of temperature exceeded 120\,000 K.  Due to the extreme temperature, the C\,\textsc{iv}, N\,\textsc{iv}\ and O\,\textsc{iv}\ lines are predicted to be too weak by the models that use the \textit{FUSE}\ temperature.  However, when looking in detail at the nitrogen lines in the spectrum of HS\,1136+6646 (Figure \ref{hs1136_line_strengths}), it is evident that neither model is able to reproduce the lines, with the N\,\textsc{iii}\ line in the model that uses the optical parameters too strong, and the N\,\textsc{iv}\ lines in the \textit{FUSE}\ parameters model too weak.

It is therefore evident that the models that use the optical parameter better reproduce the oxygen features than models that use the \textit{FUSE}\ data, while the reverse is true for other lines such as nitrogen and iron.  These differences between the models and data could be caused by a number of effects.  It is not known if either the optical or \textit{FUSE}\ stellar parameters are correct.  Therefore, if both are wrong the ability of the models to reproduce the data might be influenced by the use of the incorrect parameters.  However, this points to the underlying problem of why the optical and \textit{FUSE}\ parameters are different, and if this problem could also affect the strength of the lines in the models.  A possibility for the cause of these differences is the assumption of chemical homogeneity in the models.  Evidence of stratification of heavy elements in white dwarf atmospheres have been found by other authors \citep[e.g.][]{bars99}, and this can have a strong affect on the predicted line strengths.  For example O\,\textsc{vi}\ lines, which do not appear in homogeneous models below $\sim$50\,000 K, can be prominent at lower temperatures when a stratified model is used \citep{chay03}.  With the development of stratified models \citep[e.g.][]{schu02}, these differences might be resolved, although the assumptions made within those particular models are probably not appropriate to DAO white dwarfs, because of the processes such as mass loss that are thought to be occurring.  

A further cause of discrepancies between the data and models may be due to the limitations of the fitting process.  During the fitting procedure, the abundance of all the elements are varied by the same factor.  Therefore, the abundance of one element is measured with a model that contains the same relative abundance of all the other elements, even if those elements are not present in that particular object, or are present in different amounts.  This may alter the temperature structure of the model, changing the line strengths.  Ideally, models would be calculated for a range of abundances of every element and a simultaneous fit of all the absorption lines performed.  Such an approach may be possible in the future with increases in computing power.

In conclusion, there is disagreement between the predictions of the homogeneous models used in this work and the observed line strengths of different ionisation states when both the optical and \textit{FUSE}\ stellar parameters are used. Therefore, it is not possible to draw strong conclusions from these results to say which of the optical and \textit{FUSE}\ measurements of \teff\ are more likely to be correct.  
\section{Conclusions}
\label{conclusions}

Abundance measurements of heavy elements in the atmospheres of DAO white dwarfs have been performed using \textit{FUSE}\ data.  All the DAOs in our sample were found to contain heavy element absorption lines.  Two sets of metal abundances were determined using stellar parameters (\teff, \logg\ and $\log \frac{He}{H}$) determined from both \textit{FUSE}\ and optical data.  The results were compared to a similar analysis for DAs conducted by \citet{bars03abund}, to the predictions for radiative levitation made by \citet{chay95}, and the mass loss calculations of \citet{ungl00}.  For carbon, the DAOs were found, in general, to have higher abundances than those measured for the DAs.  When considered along with the DA abundances, the optical abundance measurements might be argued to form a trend of increasing abundance with temperature.  However, this  trend extends to a higher temperature than predicted by \citet{chay95}.  The \textit{FUSE}\ determined abundances do not follow this trend due to the higher temperatures determined from \textit{FUSE}\ data.  These abundance measurements do not increase with temperature, with no carbon abundance exceeding $10^{-4}$\ that of hydrogen, which is close to the prediction of \citet{chay95}.  The nitrogen abundances were poorly constrained for a number of the objects, and there was no obvious trend with temperature.  Although oxygen lines were identified in all the spectra, fits to those lines were, in general, very poor, making the abundances unreliable.  Silicon abundances were found to increase with temperature up to a maximum of $\sim$10$^{-4}$\ that of hydrogen, and do not show a minimum at 70\,000 K, as predicted by \citet{chay95}.  However, approximately half the fits were of poor quality, as defined by $\chi^2_{red} > $2.  Iron abundances were similar to those measured for DAs by \cite{bars03abund}\ and no trend with temperature was seen.  As with silicon, approximately half the fits were poor quality.  Nickel abundances were poorly constrained with a number of non-detections, and no trends with temperature were evident.  For none of the elements were trends with gravity or helium abundance found.  

In general, it was found that the abundances measurements that used the parameters determined from \textit{FUSE}\ data were higher than those that used the optically derived parameters by factors between 1-10.  Satisfactory fits were achieved in approximately equal number for the optical and \textit{FUSE}\ abundance measurements.  The ability of the models to reproduce the observed line strengths was slightly better for the models that used the optically derived \teff, \logg\ and $\log \frac{He}{H}$\ for oxygen and for the carbon, nitrogen and oxygen lines of those objects with \textit{FUSE}\ temperature greater than 120\,000 K.  However, the reverse is true for the iron lines.  For the example of the relative strength of N\,\textsc{iii}\ and N\,\textsc{iv}\ lines in the spectrum of one of the objects for which the optical and \textit{FUSE}\ temperatures differ, the model with the \textit{FUSE}\ temperature was better. However, for one of the objects with extreme \textit{FUSE}\ temperatures, neither model was successful at reproducing the lines.  The failure of the models to reproduce the line strengths might be caused by the assumption of chemical homogeneity.  The use of stratified models, such as those developed by \citet{schu02}\ might help to resolve the discrepancies.  In addition, the inability of the models to reproduce the lines might be influenced by the fitting technique used, in which the abundance of all elements were varied together.  In the ideal case, the abundance of all elements would be measured simultaneously.

Overall, none of the measured abundances exeeded the expected levels significantly, except for silicon.  For none of the elements, except silicon, were the abundances for the lower temperature DAOs significantly different to those for the hotter stars, despite the different evolutionary paths through which they are thought to evolve.  The abundances were also not markedly different from the abundance measurements for the DAs, apart from what might be explained by the higher temperatures.  

\section*{acknowledgments}
Based on observations made with the NASA-CNES-CSA Far Ultraviolet Spectroscopic Explorer. FUSE is operated for NASA by the Johns Hopkins University under NASA contract NAS5-32985.  SAG, MAB, MRB and PDD were supported by PPARC, UK; MRB acknowledges the support of a PPARC Advanced Fellowship.  JBH wishes to acknowledge support from NASA grants NAG5-10700 and NAG5-13213.

All of the data presented in this paper were obtained from the Multimission Archive at the Space Telescope Science Institute (MAST). STScI is operated by the Association of Universities for Research in Astronomy, Inc., under NASA contract NAS5-26555. Support for MAST for non-HST data is provided by the NASA Office of Space Science via grant NAG5-7584 and by other grants and contracts.

\label{lastpage}

\end{document}